\documentclass[twoside,agupp]{aguplus} 
\usepackage{graphicx,latexsym}
\usepackage{psfrag}
\printfigures              
\doublecaption{35pc}       
\sectionnumbers            
\lefthead{HUANG ET AL.}
\righthead{LOG-PERIODICITY IN EARTHQUAKE AFTERSHOCKS}

\received{October 28, 1999}
\revised{March 9, 2000}
\accepted{May 30, 2000}
\journalid{???}{???}
\articleid{???}{???}
\paperid{2000JB900195}
\ccc{0148-0227/00/2000JB900195\$09.00}
\cpright{AGU}{2000}

\authoraddr{Y. Huang and H. Saleur, Department of Physics, University
  of Southern California, Los Angeles, CA 90089-0484. (e-mail:
  yueqiang@usc.edu; saleur@usc.edu)}

\authoraddr{A. Johansen and M.~W. Lee, Institute of
  Geophysics and Planetary Physics, University of California, 3845
  Schlicter Hall, Los Angeles, CA 90095-1567. (e-mail:
  anders@moho.ess.ucla.edu; lee@physics.ucla.edu)}

\authoraddr{D. Sornette, Laboratoire de Physique de la Matiere
  Condensee, CNRS UMR 6622 and Universit\'e de Nice-Sophia
  Antipolis, B.P. 70, Parc Valrose, F-06108 Nice Cedex 2, France..
  (e-mail: Didier.Sornette@unice.fr)}

\newcommand \be{\begin{equation}}
\newcommand \bea{\begin{eqnarray} \nonumber }
\newcommand \ee{\end{equation}}
\newcommand \eea{\end{eqnarray}}
\newcommand{\lp}{\left(}
\newcommand{\rp}{\right)}

\slugcomment{Submitted to JGR, December 1999, JB99-0531}
\begin{document}

\title{Artifactual log-periodicity in finite size data\,:\\ Relevance
  for earthquake aftershocks}

\author{Y. Huang\altaffilmark{1,2}, A. Johansen\altaffilmark{3},
  M.~W. Lee\altaffilmark{3}, H. Saleur\altaffilmark{1}, and D. Sornette\altaffilmark{3,4,5}}




\altaffiltext{1}{Department of Physics, University of Southern
  California, Los Angeles.}
\altaffiltext{2}{Department of Earth Sciences, University of Southern
California, Los Angeles.}
\altaffiltext{3}{Institute of Geophysics and Planetary Physics, University of
  California, Los Angeles.}
\altaffiltext{4}{Department of Earth and Space Sciences, University of
  California, Los Angeles.}

\altaffiltext{5}{Laboratoire de Physique de la Matiere
  Condensee, CNRS UMR 6622 and Universit\'e de Nice-Sophia Antipolis,
  Nice, France.}

\begin{abstract}
  The recently proposed discrete scale invariance and its associated
  log-periodicity are an elaboration of the concept of scale invariance
  in which the system is scale invariant only under powers of specific
  values of the magnification factor.  We report on the discovery of a
  novel mechanism for such log-periodicity relying solely on the
  manipulation of data. This ``synthetic'' scenario for
  log-periodicity relies on two steps: (1) the fact that approximately
  logarithmic sampling in time corresponds to uniform sampling in the
  logarithm of time; and (2) a low-pass-filtering step, as occurs in
  constructing cumulative functions, in maximum likelihood
  estimations, and in de-trending, reddens the noise and, in a finite
  sample, creates a maximum in the spectrum leading to a most probable
  frequency in the logarithm of time.  We explore in detail this
  mechanism and present extensive numerical simulations.  We use this
  insight to analyze the 27 best aftershock sequences studied by
  \citet{kisslinger199107} to search for traces of genuine
  log-periodic corrections to Omori's law, which states that the
  earthquake rate decays approximately as the inverse of the time
  since the last main shock.  The observed log-periodicity is shown to
  almost entirely result from the ``synthetic scenario'' owing to the
  data analysis. From a statistical point of view, resolving the issue
  of the possible existence of log-periodicity in aftershocks will be
  very difficult as Omori's law describes a point process with a
  uniform sampling in the logarithm of the time. By construction,
  strong log-periodic fluctuations are thus created by this
  logarithmic sampling.
\end{abstract}

\begin{article}

\section{Introduction}\label{intro}

Log-periodic corrections to scaling have been one of the most exciting
topics in the physics of rupture in recent years
\citep{saleur199603,saleur199603a,saleur199608}.  If present, these
corrections might make the prediction of major events a less hopeless
task \citep{anifrani199506} than previously thought \citep{geller97}.
Log-periodicity is a characteristic behavior of systems exhibiting
discrete scale invariance (DSI) \citep{sornette1998pr}, either
geometrical or spontaneously generated by the dynamics. Like all
symmetries, DSI could provide a powerful new concept in the study of
cracks, rupture, and disordered systems in general.

DSI is an elaboration of the concept of scale invariance, which has
been broadly applied in the geosciences
\citep{scholz89:_fract,barton95,turcotte97,sorbook}.  The word
``fractal'' was coined by B.B. Mandelbrot to describe sets which consist of
parts similar to the whole and which can be described by a fractional
dimension. This generalization of the notion of a dimension from
integers to real numbers reflects the conceptual jump from translation
invariance to continuous scale invariance. More recently, the
generalization of the notion of dimension, according to which the
dimensions or exponents are taken from the set of complex numbers, has
been shown to describe the interesting and rich phenomenology of
systems exhibiting discrete scale invariance, associated with
log-periodic corrections to scaling
\citep{saleur199603,saleur199603a,saleur199608,sornette1998pr}.
Discrete scale invariance is a weaker kind of scale invariance in
which the system or the observable obeys scale invariance only for
specific choices of magnifications, which form, in general, a countably
infinite set of values, all powers of a fundamental scale factor
$\lambda$.  This property can also be seen to include the concept of
lacunarity of the fractal structure.

Log-periodic oscillations have been reported in acoustic emissions
prior to rupture
\citep{anifrani199506,sahimi9610,johansen199805,sornette99}, to rock
bursts in deep mines \citep{ouillon00}, in precursory seismic activity
before large earthquakes \citep{sornette199505}, and in anomalous
ionic concentration prior to the Kobe earthquake
\citep{johansen199610,johansen99:_new_kobe}. The status of
log-periodicity is the clearest for rupture of heterogeneous materials
for which a sub-harmonic cascade of Mullins-Sekerka instabilities has
been recognized as the probable underlying mechanism
\citep{sornette199601a,huang199706}. The application to earthquakes is
controversial \citep{gross199804} as only a few case studies have been
reported. In this context, a major difficulty consists of the
selection of the appropriate space-time window to construct
time-to-failure functions \citep{bowman199810,brehm99}. Another
problem stems from the scarcity of seismic data relevant to this
problem since only large (and thus rare) earthquakes exhibit
precursory seismic activity observable solely on relatively large
foreshocks \citep{reasenberg99}.  Notwithstanding these limitations,
further investigation of the possible existence of log-periodicity is
important for its potential use (1) in constraining the mechanisms
controlling earthquake triggering, (2) in focusing possible
forecasting skills based on the measured accelerated seismicity, and
(3) in providing indirect measurements of the stress drop ratio and
the seismic efficiency \citep{leephd}.

It is fair to say, however, that the existence of log-periodic
corrections in rupture is not proven beyond doubt: They have typically
been observed in systems with very noisy data, and so far, we are
still lacking a nontrivial analytically solvable model where they
could be shown to exist rigorously. By nontrivial, we mean a realistic
model of rupture without preexisting hierarchy which could be treated
analytically to prove log-periodicity.  In that respect, the situation
is not unlike critical phenomena, where the possibility of power law
singularities was hotly debated before \citet{onsager44}'s solution of
the Ising model.  This must be contrasted with the preexisting
hierarchical geometry, such as the hierarchical diamond lattices or
tent-like hierarchies or Cantor sets used by \citet{newman9511} and
\citet{saleur199603,saleur199603a,saleur199608} for which the
log-periodicity derives rather trivially by measuring the fractal
dimensions of the underlying geometrical object
\citep{sornette1998pr}.

It is therefore of vital importance to have a greater understanding of
the relation between apparent log-periodic corrections and noise.
After the initial period of enthusiasm, the suspicion has grown, in
particular, that the consideration of cumulative quantities, like the
Benioff strain in foreshock sequences, might well introduce spurious
log-periodicity in the data, despite the lack of any physical DSI.
The purpose of this paper is to examine several scenarios where
apparent log-periodic corrections are generated by the interplay of
noise and the method of data analysis.  We report on a basic
artifactual source generating log-periodicity, which is a combination
of a sampling that is periodic in a log timescale and some type of
low-pass filtering such as in constructing cumulative functions,
maximum likelihood estimators, and de-trending.

In the analysis of power law data, sampling is more often regular
(sometimes even periodic) on a log scale than might be initially
expected.  For instance, in our previous analysis of the Kobe
earthquake the chemical concentrations reported by
\citet{tsunogai95,tsunogai96} were sampled in a nonuniform way
because the measurements were performed {\bf after} the earthquake.
Specifically, the chemical composition of groundwater prior to the
earthquake was analyzed by collecting bottles of mineral water from
wells close to the epicenter for which the production dates could be
identified. Since the depletion process of bottles in stores can be
approximated by a Poisson process with some rate $\mu$, this implies
that the number of bottles with a production date $t$ prior to the
date of the earthquake $t_c$ is proportional to $\exp [ - \mu \lp t_c
- t\rp]$, quantifying an accelerated sampling in the time $t_c - t$ to
the earthquake. This is not exactly a sampling that is uniform in the
variable $\ln \lp t_c - t \rp$; it will nevertheless be approximately
so in finite time intervals and, as a consequence, may lead to the
effects discussed in this paper.

Another example occurs naturally when analyzing aftershocks.  Omori's
law for aftershocks \citep{utsu199501} states that the rate $Q(t)$ of
earthquakes following a main earthquake decays as $t^{-p}$, where $t$
is the time from the main shock and $p$ is close to $1$:
\be
Q(t) \propto 1/t^p\,.         \label{omorilaw}
\ee
It is equivalent to having the probability of an aftershock in the
time interval $[t, t+dt]$ proportional to $dt/t^{p}$. Alternatively,
the occurrence time $t$ for an aftershock is a random variable with a
power law probability distribution.  When analyzing a given empirical
aftershock time series, the problem of measuring the rate precisely
becomes crucial.  To compute the rate, one has to choose a
binning interval. A constant interval is not satisfactory since
Omori's power law implies that events cluster densely at the beginning
of the time series and become very sparse at long times.  A binning
interval varying with time is the most natural, but it will, when
optimally chosen, realize once again a periodic sampling in log scale.
We will discuss this in great detail in Section \ref{logbinning}.

The spurious log-periodicity generated by such sampling will usually
become amplified when considerating cumulative quantities, which act
as a low-pass filter. This is quite unfortunate, since taking
cumulative quantities is a very natural procedure in many cases.  For
instance, taking the cumulative of the empirical rate of aftershocks
appears to be a natural well-defined approach, which has the advantage
of being nonparametric and amounts simply to counting the number of
events up to a running time $t$.

The paper is organized as follows. In Section \ref{cumulsect} we
analyze in detail the effect of cumulating quantities on
log-periodicity. In Section \ref{create} we show that when the
binning of a power law is logarithmic in time, the cumulative
distribution creates splendid and totally spurious log-periodicity.
Section \ref{ns} applies this insight to real aftershock sequences:
Using this cumulative method, we find that it is not possible to
distinguish any real log-periodicity from the synthetic mechanism in
the 27 best aftershock sequences studied by \citet{kisslinger199107}
and in the Loma Prieta, Landers, and Northridge aftershock sequences.
Section \ref{sec:tests-real-aftersh} shows that, conversely,
cumulating can destroy a preexisting log-periodicity, when the
binning is performed regularly and nonlogarithmically.  Section
\ref{logbinning} turns to another analyzing method using maximum
likelihood. We again show that the synthetic mechanism for
log-periodicity operating in the cumulative method is present
here, and we quantitatively demonstrate its main properties. Applied to
this set of aftershock sequences, we find that most of the observed
log-periodic signal can be explained by the synthetic mechanism.
However, there seems to remain a residual signature, possibly stemming
from a physically based log-periodicity, which can not be accounted
for by the synthetic mechanism.  This evidence is not sufficiently
strong to conclude definitively with high statistical significance.
Section \ref{sec:discussion} summarizes our findings.

\section{Cumulative Distributions and Their Fluctuations}
\label{cumulsect}

Omori's law (\ref{omorilaw}) and, more generally, any continuous law can be
obtained from a one-to-one mapping to a uniformly distributed random
variable $x$ (without loss of generality, we take $x \in [0,1]$ when
not specified otherwise):
\begin{equation}
  \label{pxpt}
  x=\int_0^t Q(t')dt'\,,
\end{equation}
which is the cumulative distribution function (cdf) obtained from the
condition of ``conservation of probability'' $P(x)dx=Q(t)dt$ under a
change of variable.  Consider the special case $p=1$. This leads to
\begin{equation}
  P(x)dx \sim dx \sim \frac{1}{t} dt \sim d\ln t\,,   \label{rqjijka}
\end{equation}
which highlights the mapping between the linear scale in $x$ and the
logarithmic scale in $t$.

\begin{figure}
  \psfrag{\(a\)}[][]{(a)}
  \psfrag{(b)}[][]{(b)}
  \psfrag{t}[][]{$t$}
  \psfrag{t0}[][]{$t_0$}
  \psfrag{x}[][]{$x$}
  \psfrag{x0}[][]{$x_0$}
  \psfrag{log t}[][]{$\ln(t)$}
  \figbox*{\hsize}{}{\includegraphics[width=8cm]{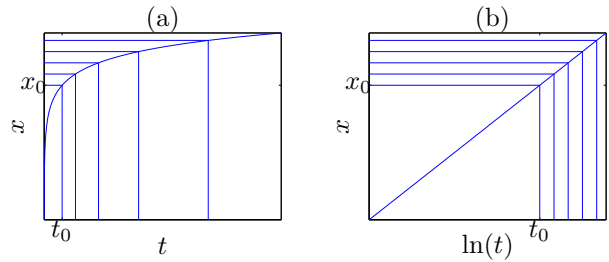}}
  \caption{Mapping between a uniformly distributed random
        variable $x$ and a random variable $t$ with power law
        distribution: (a) linear scale for $t$ and (b) log scale for $t$.}
  \label{mapping}
\end{figure}

The mapping (\ref{pxpt}) is shown graphically in Figure \ref{mapping}.
A given value $x_0$ on the $x$ axis corresponds to a point $t_0$ on
the $t$ axis, through the transformation (\ref{pxpt}) represented by
the continuous curve in Figure \ref{mapping}a.  If $x_0$ is a random
variable uniformly distributed on the $x$ axis, $t_0$ will be a random
variable with a power law distribution. A sequence of equally spaced
$x$ is mapped onto a sequence of $t$ with increasing separation
following a power law $p(t)$. Note that we are not sampling a power
law function, rather the power law probability density function (pdf)
results from the spacing between the times $t$ of the aftershocks.
When $t$ is plotted on a log scale (Figure \ref{mapping}b), we see
that if there is a cluster of points near $x_0$, there will be a
cluster of points near $t_0$.  When the clustering of points in $x$ is
periodic on the $x$ axis, the clustering is periodic on the $\ln
t$ axis; thus periodicity in the cdf of $x$ is mapped to
log-periodicity in the cdf of $t$.

\subsection{Theory\label{noise}}

\subsubsection{Shot-noise spectrum for cumulative distributions of finite
data sets.}

To understand the effect of constructing cumulative quantities, we
consider the cdf for the variable $x$ uniformly distributed in
$[0,u]$. The cdf is by definition cdf$(x)=x/u$ defined for $0 \leq x
\leq u$.  Here we add the interval size variable $u$ to investigate
how the fluctuations, in particular the most probable ones, may depend
on the size of the sampled interval.  Let us assume that a finite
number $N$ of values $x$ are sampled in the interval $[0,u]$, and let
us construct the empirical cdf cdf$_e(x)$ from this data set by simply
counting the number of values less or equal to $x$ and normalizing by
$N$. We expect to get an approximate straight line with step-like
fluctuations around it. These ``shot noise'' fluctuations are
described by
\be
{\rm cdf}_e(x) = {1 \over N} \sum_{i=1}^N H(x-x_i)~,
\ee
where the Heaviside function: $H(y) = 0$ for $y<0$ and is $1$
otherwise.  To study the fluctuations around the theoretical
cdf$(x)=x/u$, we estimate the Fourier transform of the residual $R
\equiv$ cdf$_e(x) - x/u$, which reads
\be
R = - {1 \over i \omega} \biggl[ {1 \over N} \sum_{i=1}^N  e^{i\omega x_i}
~~-~~ {e^{i\omega u} - 1 \over i \omega u} \biggl]~.   \label{mgfkjakkf}
\ee
The first term on the right-hand side gives the usual fluctuation spectrum
$|R|^2 = N^{-1} \omega^{-2}$ of ``red noise.'' The other term in the
r.h.s. is only present for {\it finite} interval size $u$. In the
presence of a finite $u$, the spectrum $|R|^2$ presents correction
factors with amplitude proportional to $(e^{i\omega u} - 1)/i
  \omega u$ and its modulus squared. These corrections describe the
possible existence of oscillations with angular frequency $\omega$
decorating the linear dependence.  This shows us that the finite size
$u$ of the interval introduced a nonzero most probable angular
frequency. Here, by most probable, we mean the frequency corresponding
to the largest peak of the spectrum.  It is defined as the angular
frequency that maximizes
\be
\Bigg|{e^{i\omega u} - 1 \over i \omega u}\Bigg|^2 = \frac{2}{\omega^2} [1-\cos(\omega u)]~,
\ee
which is the solution of
\be
\tan(\omega u/2) = \omega u/2~.
\ee
Thus we find $\omega u/2 \approx 3\pi/2$.  Since $\omega=2\pi f$, this
gives the most probable frequency
\be
f \approx 1.5/u~, \label{jmmafqqmq}
\ee
which is inversely proportional to the length $u$ of the interval.
This constitutes our first prediction.

From the structure of (\ref{mgfkjakkf}), we see that the amplitude of
the leading correction to asymptotic spectrum $N^{-1} \omega^{-2}$ is
proportional to $1/N$; that is,  we expect the most probable oscillation
decorating the linear cdf $x/u$ to have a typical amplitude
proportional to $1/\sqrt{N}$. This constitutes our second prediction.

\subsubsection{Spectrum of cumulative variables that are solution of a
stochastic ordinary differential equations.}

To show the generality of these results, we can generalize the problem
within the framework of stochastic ordinary differential equations and
study the ``cumulative'' variable $X$ defined by
\be
dX/du = a X + \eta(u) ~,         \label{eq1}
\ee
where $\eta$ is a white noise defined by $\langle \eta(u) \eta(u')
\rangle = \delta(u-u')$, where the angle brackets mean that an ensemble
average is taken.  Eq. (\ref{eq1}) allows for a deviation from
the simple linear cumulative law by introducing a first-order
autoregressive correlation (the $aX$ term). The solution of
(\ref{eq1}) is
\be
X(u) = \int_0^u \eta(\tau) \exp[a(u-\tau)] d \tau~.  \label{eq2}
\ee
Eq. (\ref{eq2}) shows that $X$ is the cumulative of $\eta$,
weighted by the exponential term $\exp[a(u-\tau)]$ (equal to $1$ for
$a=0$).  The white noise can be written as
\be
\eta(\tau) = \int d\omega ~e^{i \omega \tau} {\hat \eta}(\omega)   ~,
\label{eq3}
\ee
which defines ${\hat \eta}(\omega)$. The stochastic variable is
characterized by the following covariance: $\langle {\hat
  \eta}(\omega){\hat \eta}^*(\omega') \rangle =
\delta(\omega-\omega')$, where the asterisk denotes the complex conjugate.
Putting (\ref{eq3}) into (\ref{eq2}), we get
\be
X(u) = \int d\omega~ f_u(\omega)  e^{i \phi(\omega)} ~,
\ee
where
\be
f_u(\omega) = {e^{i\omega u} - e^{a u} \over i\omega - a}~.   \label{hqk}
\ee
The most probable angular frequency is again obtained by maximizing
$|f_u(\omega)|^2$.  Let us take the limit of small $a$ for which the
noise dominates (in the power law mapping, this corresponds to an
exponent $p$ close to $1$). Then (\ref{hqk}) simplifies into
\be
f_u(\omega) \approx {e^{iu \omega} - 1 \over i\omega}
\ee
\be
|f_u(\omega)|^2 = (2/\omega^2) [1-\cos(\omega u)]  ~. \label{prediajjak}
\ee
This retrieves the same solution (\ref{jmmafqqmq}) as above for the
shot noise problem.  Note that the $1/\omega^2$ spectrum for a signal
with an infinite range is modified by the cosine correction observed
for a finite range.

These simple calculations can be explained intuitively as follows.
Consider an uncorrelated Gaussian noise. In the frequency domain the
spectrum of Gaussian noise is flat (white), since all frequency
components are equally probable.  Let us call $G(\omega)$ the white
spectrum of the measurement error $g(x)$.  $G(\omega)$ does not,
however, extend to all frequencies and is bounded by the measurement
range of $x$.  The lowest angular frequency $\omega_{\mathrm{min}}$ is
determined by the span of $x$, corresponding to one half cycle
covering the whole interval (Nyquist frequency). The highest angular
frequency is determined by the distance between successive $x_i$.

When noise is integrated, as occurs in constructing a cumulative
distribution, the resulting noise is smoother and its spectrum is
reddened with more power in the low-frequency band. This results from
the fact that the spectrum of $\int^x g(x')dx'$ is
$\omega^{-2}~G(\omega)$ \citep{strat67}.

Regardless of the precise shape of $G(\omega)$ (we allow here for
departure from white noise), the combination of the lower angular
frequency cutoff and the $\omega^{-2}$ increase implies that
$\omega^{-2} G(\omega)$ presents a well-defined peak near
$\omega_{\mathrm{min}}$. The left side of the peak is determined by
the nature of the frequency cutoff, while the right side is controlled
by the $\omega^{-2}$ decay resulting from the integration. This
angular frequency peak corresponds to the most probable oscillation in
the signal, \emph{i.e.}, the frequency at which the spectrum is maximum. In
other words, the integration of the white noise in finite time series
creates a low-frequency periodic oscillation decorating the average
trend of $\int^x g(x') dx'$.

We thus obtain the following general result: for approximately
uniformly distributed random variables and finite samples, the most
probable oscillation decorating the cumulative distribution is
inversely proportional to the range $u$ of the variables. In addition,
the amplitudes of the oscillations decay on average as the square root
of the sample size.  This is therefore a finite size effect.

\citet{slutzky37} similarly found that taking sums {\bf and}
differences of discrete time series, \emph{i.e.}, filtering-smoothing them,
introduces oscillations in otherwise completely noisy data. See also
\citet{zaj76} for simulations with an application to economy.

\subsubsection{Application to log-periodicity in cumulative power law
distributions.}
\label{create}

With the change of variable $u \to \ln t$ obtained in (\ref{rqjijka}),
for $p=1$, (\ref{eq1}) describes a power law $X \sim t^a$ decorated
by noise. The previous calculations in the linear variable $u$
translates directly to the variable $\ln t$, predicting a most
probable log-periodic component with log frequency $1.5/\ln(t_2/t_1)$,
where $[t_1, t_2]$ defines the interval in the variable $t$.

\begin{figure}
  \psfrag{log\(t\)}[][]{$\ln(t)$}
  \figbox*{\hsize}{}{\includegraphics[width=8cm]{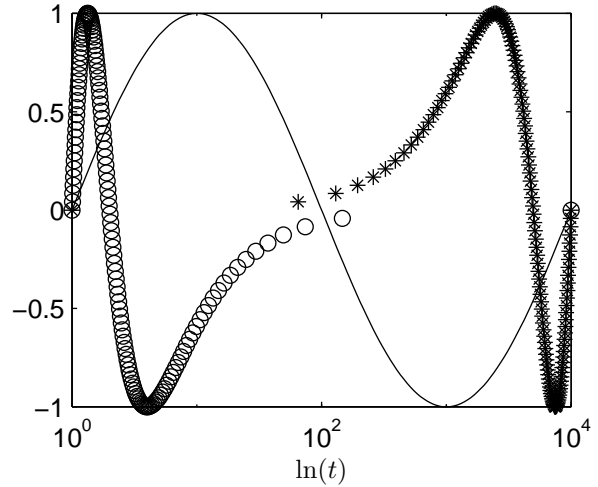}}
  \caption{Mapping of a periodic function in $x$ under the
      transformation (\ref{pxpt}) for $p=1$ (solid line), $p=2$
      (circles), and $p=0.01$ (stars). }
  \label{mne1}
\end{figure}

When the power exponent $p$ is not exactly $1$, the mapping curve in
Figure \ref{mapping}b will not be a straight line, and then the mapping
of a periodic function in $x$ under the transformation (\ref{pxpt})
will not yield an exact periodic function, as shown in Figure
\ref{mne1}.

In Figure \ref{mne1} when the power law exponent $p$ is $1$ (solid
line), the function is periodic in $\ln t$. For $p=2$ (circles) the
function is not exactly periodic in $\ln t$. Owing to the rapid decay of
$1/t^2$, the events cluster closer to $t=0$ than for the $1/t$ decay.
In contrast, for $p<1$ the events cluster at larger $t$ more than for
the $1/t$ decay, which leads to the oscillations shown by the stars in
Figure \ref{mne1} for $p=0.01$. For more reasonable values closer to
$1$, we observe significant regular oscillations which are
approximately log-periodic. Their significance is assessed by the
existence of a significant peak in the spectrum. The approximate
log frequency peak will be higher in the case of $p \neq 1$ than in
the case for $p=1$, but the difference is not very strong for the
empirically relevant cases for which the apparent exponent does not
deviate very much from $1$. In the case of aftershocks, Omori's law is
typically characterized by exponents in the range $[0.5, 2.0]$
\citep{kisslinger199107}.  Similarly, the accelerating Benioff strain
of precursory earthquake activity prior to a main shock is typically
characterized by a power law $(t_c-t)^{-p}$ with an exponent $p$ close
to $0.5$ \citep{bowman199810}.

Therefore the generation of log-periodicity in the case of
logarithmically sampled power laws relies on the following
ingredients\,: (1) noise and sampling fluctuations (shot noise)
provide all kinds of frequency components; (2) integration filters out
high-frequency components, enhances low-frequency components, and
produces a $\omega^{-2}$ spectrum only perturbed for the lowest
frequencies for the finite range in $\ln t$ of the data set; and (3) there
is a direct mapping from periodicity to log-periodicity for power laws
with exponents close to $p=1$ since data points are close to being
uniformly distributed in $\ln t$ scale.

Summarizing, we expect to observe log-periodic oscillations from the
integration of noise in power laws\,:
(1) the (log-)frequencies of the log-periodic oscillations should
  have a power spectrum given by $\omega^{-2}$;
(2) the most probable frequency should scale as the inverse of the
  range in log scale of the analysis;
(3) the most probable frequency should increase when the power law
  exponent $p$ deviate from $1$;
and (4) the amplitude of the oscillations should scale as the inverse
  square root of the number of data points (central limit theorem).

In Section \ref{ns} we present numerical tests of these predictions
performed on synthetically generated samples.

\subsection{Numerical Simulations}
\label{ns}

Our numerical simulations are based on the hypothesis that the noise
is uncorrelated. Empirical realizations of the noise in real data sets
may be of a different nature.  However, we believe that our results
hold by and large as our mechanism relies mainly on the existence of
an increasing power spectrum at low log frequencies bounded from below
by finite size effects.

\begin{figure}
  \figbox*{\hsize}{}{\includegraphics[width=8cm,bb=62 398 310 604,clip]{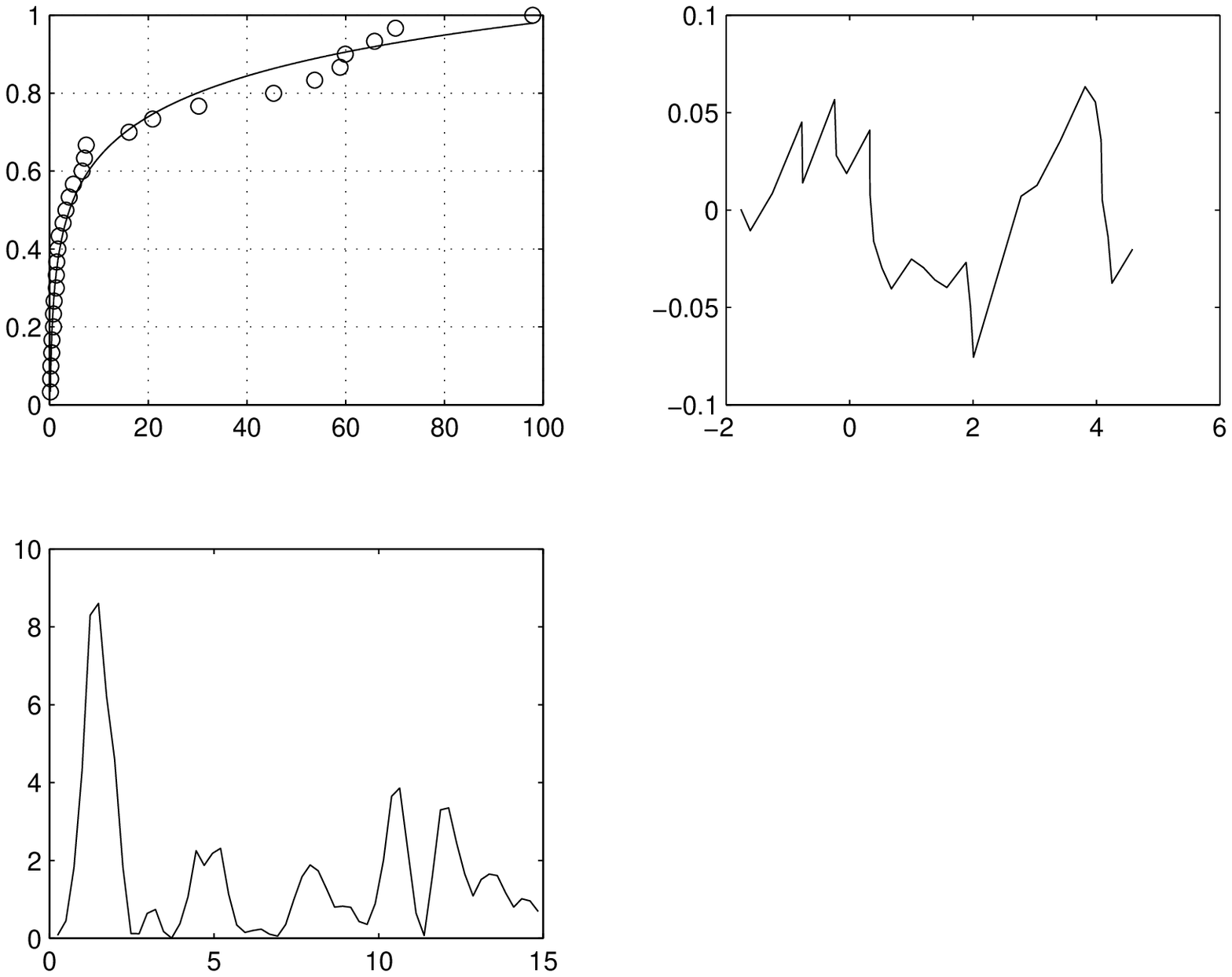}}
  \caption{Circles represent the
      cumulative distribution of a realization of synthetic
      aftershocks with Omori's law ($p=1$) obtained with (\ref{t}). The
      solid line is a fit to the synthetic data using a power law. The
      vertical axis is the normalized cumulative number of events. The
      horizontal axis is time $t$ (the same unit as the real sequence).}
  \label{aftshkexample1a}
\end{figure}

A sequence of event occurrence times $t_i$ for $i=1,..,N$ are
generated within the fixed time interval $[t_0, t_u]$, with a power
law rate $1/t^p$ by the transformation method \citep{press1992}.  In
this method, one starts from a uniformly distributed random variable
$x \in [0,1]$ and generates the corresponding random variables $t$
using the transformation (\ref{pxpt}), \emph{i.e.},
\begin{equation}
  \label{t}
  t=\left[ t_0^{1-p} + x \left( t_u^{1-p} - t_0^{1-p} \right)
\right]^{\frac{1}{1-p}}\,,
  \label{hgahaka}
\end{equation}
on each of the realizations of the $x$ variable. In this procedure we
specify the number $N$ of points $x$ in $[0,1]$, generate them in this
interval using a random number generator, and apply the transformation
(\ref{hgahaka}) to each of these values to obtain the sequence $t_i$
for $i=1,..,N$.  The cumulative distribution of a typical realization
(circles) is shown in Figure \ref{aftshkexample1a}.

\begin{figure}
  \figbox*{\hsize}{}{\includegraphics[width=8cm,bb=312 406 544 589,clip]{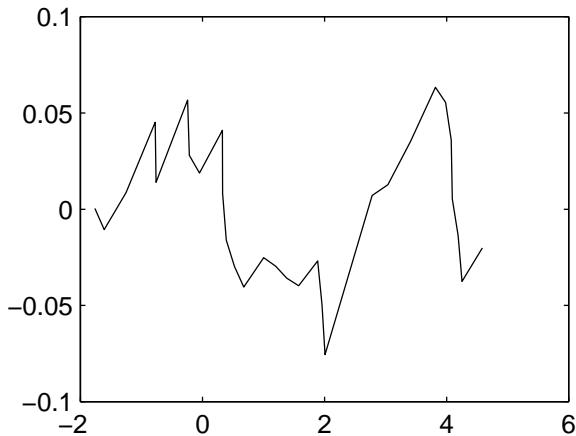}}
  \caption{Oscillations around the power law (solid line) of the
      normalized cumulative number of events (circles) in Figure
      \ref{aftshkexample1a}. The horizontal axis is $\ln t$.}
  \label{aftshkexample1b}
\end{figure}

In our analysis, we first determine the best power law fit to the
synthetic data of Figure \ref{aftshkexample1a} and study its residue,
shown in Figure \ref{aftshkexample1b}.  We then perform a spectral
analysis of the residue to test for the presence of possible periodic
oscillations. Since the data points are not evenly spaced in the $\ln
t$ scale, we use the Lomb method \citep{press1992}.  Recall that for
unevenly sampled data the Lomb method is superior to fast Fourier
transform (FFT) methods because it weights data on a ``per point''
basis instead of a ``per time interval'' basis. Furthermore, the
significance level of any frequency component can be estimated quite
accurately \citep{press1992}.

%

\begin{figure}
  \figbox*{\hsize}{}{\includegraphics[width=8cm,bb=71 211 312 404,clip]{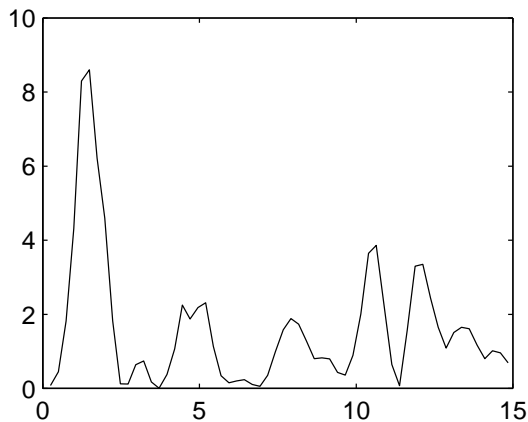}}
  \caption{The spectrum of the residue
    shown in Figure \ref{aftshkexample1b}. The horizontal axis is
    frequency $\omega$. The vertical axis is the normalized Lomb
    spectral power.}
  \label{aftshkexample1c}
\end{figure}

In order to determine the presence or absence of a periodic signal in
the residue, it is necessary to quantify the statistical significance
of the peaks found in the spectrum $P_{N}(\omega)$.  We use the null
hypothesis that the noise realizations at distinct times are
uncorrelated Gaussian distributed, which in the Lomb method leads to
$P_{N}(\omega)$ with an exponential probability distribution with unit
mean. Thus, if we scan $M$ independent frequencies, the probability
that none has a spectral peak larger than $z$ is $(1-e^{-z})^M$. The
false alarm probability of the null hypothesis is then
\begin{equation}
  \label{sig}
  P({\rm peak}>z)\equiv 1-(1-e^{-z})^M\,.
\end{equation}
A small value for the false alarm level indicates a highly significant
periodic signal. In general, $M$ depends on the number of frequencies
sampled, the number of data points, and their detailed spacing.
Usually, $M$ is of the order of a few $N$, the number of data points
(for example in the Interactive Data Language (IDL) implementation of
the Lomb method, $M=N$).  Figure \ref{aftshkexample1c} shows the
spectrum of the residue presented in Figure \ref{aftshkexample1b}. One
observes a main peak near $\omega = 1.5$, with height (with normalized
spectral density) around $8.5$.  Since the number of data points here
is $N=31$, using the standard choice $M=N=31$ gives the false alarm
level $0.006$ for this peak, indicating that the probability that this
peak results from chance is 0.6\%, which corresponds to a confidence
level better than 99\%.

\begin{figure}
  \psfrag{w}[][]{$\omega$}
  \psfrag{N}[][]{$N(> \omega)$}
  \psfrag{slope -1.2}[][]{slope -1.2}
  \figbox*{\hsize}{}{\includegraphics[width=8cm]{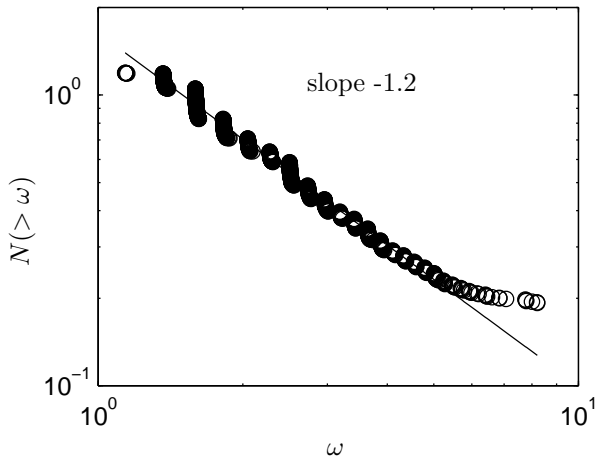}}
  \caption{The cumulative distribution of the most significant
    angular frequencies of the log-periodic oscillations obtained from
    the cumulative power law distribution.  The straight line is the
    theoretical prediction of $\omega^{-2}$ (corresponding to slope -1
    in the cumulative distribution) and is included to guide
    the eye.}
  \label{wdis}
\end{figure}

In the following, we will consider only the most significant peak of
our synthetic samples.  We first run 1000 simulations with $N=100$
data points with $t_0=0.1$ and $t_u=100$.  Using the estimation of the
false alarm probability (\ref{sig}), we find that 99.4\% among the
1000 synthetic time series have a significance level better than 99\%
for the presence of a periodic component.  The distribution of these
angular frequencies is shown in Figure \ref{wdis} and is compatible
with the theoretical prediction $\sim \omega^{-2}$ given by
(\ref{prediajjak}).

\begin{figure}
  \psfrag{1/log\(t range\)}[][]{$1/[\ln(t_{\mathrm{max}}/t_{\mathrm{min}})]$}
  \psfrag{frequency}[][]{Frequency $\omega$}
  \figbox*{\hsize}{}{\includegraphics[width=8cm]{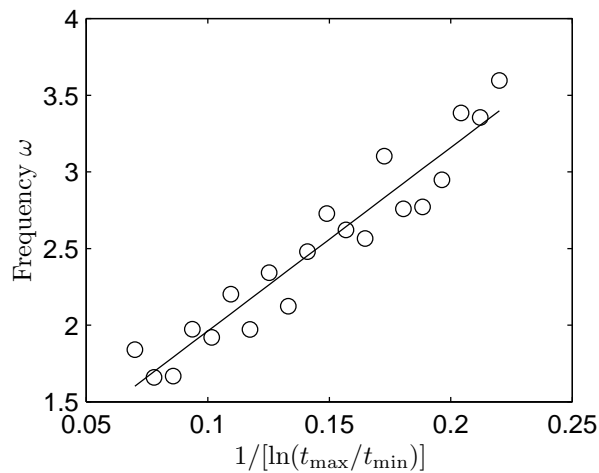}}
  \caption{The average angular frequency as a function of
        $1/[\ln(t_{\mathrm{max}}/t_{\mathrm{min}})]$.}
  \label{wtrange}
\end{figure}

We now test the prediction (\ref{jmmafqqmq}) that the most probable
angular frequency is inversely proportional to the range
$\ln(t_{\mathrm{max}}/t_{\mathrm{min}})$ in $\ln t$ of the data set.
In our tests, we set $t_{\mathrm{min}}=1$ and sweep
$t_{\mathrm{max}}$.  For each value of $t_{\mathrm{max}}$ we
construct 100 runs to get the average angular frequency. Since the
angular frequencies are distributed according to a power law
(approximately $\omega^{-2}$), the average is not the most probable
value. However, both quantities have very similar distributions, and there
is a simple proportionality relation between the average and the most
probable value.  We analyze the average angular frequency for
computational simplicity.  Figure \ref{wtrange} shows the average
angular frequency as a linear function of
$1/[\ln(t_{\mathrm{max}}/t_{\mathrm{min}})]$.
%

\begin{figure}
  \psfrag{m}[][]{$m$}
  \psfrag{w}[][]{$\omega$}
  \figbox*{\hsize}{}{\includegraphics[width=8cm]{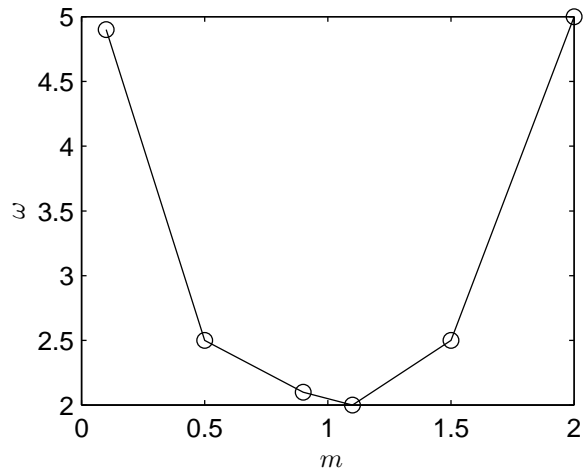}}
  \caption{The most probable angular frequency $\omega$ as a function
    of the power law exponent $m$.}
  \label{wm}
\end{figure}

As seen in Figure \ref{mne1}, when the power law exponent $p$ is not
equal to $1$, the mapping between periodicity in the $x$ and
log-periodicity in the $t$ variable is not exact\,: the clustering of
data points either at small $t$ ($p>1$) or at large $t$ ($p<1$)
effectively shortens the logarithmic range and thus increases the
effective value of the most probable frequency. We verify this
prediction by generating, for each value of $p$, 20 runs that allow
us to get an estimation of the log frequency distribution from which
the most probable log angular frequency is extracted. We verify in
Figure \ref{wm} that the most probable log angular frequency increases
as $p$ departs from $1$.

\begin{figure}
  \psfrag{amplitude}[][]{Amplitude}
  \psfrag{N}[][]{$N$}
  \figbox*{\hsize}{}{\includegraphics[width=8cm]{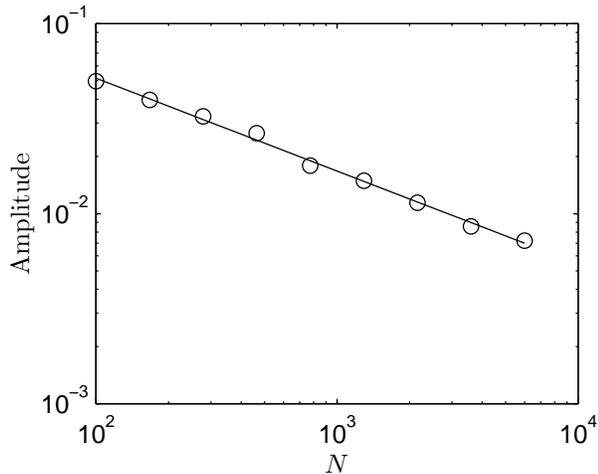}}
  \caption{The scaling relation between the amplitudes of the log-periodic
      oscillations and the number of data points $N$. The slope is $-0.49$.}
  \label{amp}
\end{figure}

We now test the prediction that the amplitude of the log-periodic
oscillation should be proportional to $N^{-1/2}$, where $N$ is the
number of data points. For each value of $N$ the amplitude is
averaged over 20 runs. Figure \ref{amp} shows the dependence of the
log-periodic oscillation amplitude for $N$ going from 100 to 5995,
with the best fit $\sim N^{-0.49}$ in excellent agreement with the
prediction.

Although log-periodic oscillations from integration of noise tend to
be low log frequency, corresponding to one to two oscillation cycles, our
simulations show that the probability of observing oscillations that
span up to 2.5 periods is still around 10\%.

\subsection{Tests on Real Aftershock Data}
\label{sec:tests-real-aftersh}

Recent numerical simulations and a mean-field analysis have been
performed on models of earthquake aftershocks in one-, two-, and
three-dimensional Euclidean lattices, using both continuous elasticity
and discrete cellular automata modeling \citep{leephd,lee9903}. Owing
to the interplay of power law stress corrosion and the threshold
nature of rupture, the average stress is found to decay in a
punctuated fashion after a main shock, with events occurring at
characteristic times increasing as a geometrical series with a
well-defined multiplicative factor (a signature of log-periodicity)
which is a function of the stress corrosion exponent, the stress drop
ratio, and the degree of dissipation.  These results, found independent
of the discrete nature of the lattice, raise the possibility that
log-periodic structures may also be observable in real aftershock data
\citep{leephd}.  If confirmed, their existence would provide
constraints on their underlying mechanism as well as on the stress
corrosion exponent, the stress drop ratio, and the degree of
dissipation.  Another possible mechanism comes from the hierarchical
structure of the fracturation of the crust with the coexistence of an
approximate self-similarity together with characteristic scales
associated with the thicknesses of geological and rheological layers
\citep{ouillon95,ouillon96}.  These considerations have motivated us
to search for log-periodic oscillations around Omori's power law of
the normalized cumulative number of events for 27 aftershock sequences
in southern California, 1933-1988.  The log-periodic components in the
oscillations are extracted using the Lomb method and compared with
those from synthetic aftershock sequences generated with the same
parameters as the real aftershock sequences.

In order to avoid possible bias in the selection of the aftershock
time series we decided to study a sufficiently large and high-quality
data set that has been studied previously by an independent team,
namely, the 39 aftershock sequences in southern California covering
the time from 1933 to 1988, compiled by \citet{kisslinger199107} using
the \citet{reasenberg198506} algorithm.  In addition, we studied more
recent large aftershock time series from the Loma Prieta earthquake
(October 1989 in northern California), the Landers earthquake (June 1992),
and the Northridge earthquake (January 1994).

\citet{kisslinger199107} used these sequences to study the properties
of aftershocks in southern California, focusing on Omori's power law
$1/t^p$ and possible correlation between the exponent $p$ and
geophysical properties like surface heat flow.  They found that the
temporal behavior of aftershocks is generally well-described by the
modified Omori's law.  The $p$ values range from 0.69 to 1.809, with
mean value of $1.109$ and standard deviation $0.246$.  Among the
sequences, six values were graded ``poor,'' 10 ``fair,'' and 24
``good,'', on the basis of the relative size of the standard error to the
$p$ value itself with a standard error $<0.1\ p$ being
good, $>0.2\ p$ being poor, and fair being in between.
The number of aftershocks in one sequence ranges from 12 to 1437, and
the duration of the sequences ranges from 7 days to around 1000 days.
Note that the sizes of the samples are adjusted by the chosen magnitude
cutoffs which have been varied to test the robustness of our results.

In order to test for the possible existence of log-periodic
oscillations decorating Omori's power law we need the $1/t^p$ power
law to be a good approximation in the first place. We thus discarded
the poor data sets (4, 7, 8, 26, 34, 35) as defined by
\citet{kisslinger199107}.  We then examined the other sequences very
carefully before using them, and some more sequences were discarded for
some obvious reasons: 

1. A sequence is severely incomplete in the early stage as seen from
the plot of the temporal distribution of the events (4, 7, etc.,
confirming Kisslinger and Jones' fits).

2. There is a sudden change of properties in the middle of the
sequence due to some large aftershocks which probably caused their own
aftershocks (19, 21, 37, 38). This is explicitly illustrated for the
two sequences 19 and 37 (for sequence 19, $p=5.8 \pm 9.0$ from a
fitting program in a commercial software system Matlab using the
Levenberg-Marquardt algorithm, while Kisslinger and Jones' result is
$p=1.113\pm0.415$.  For other sequences, our results are close to
their results. For sequence 37, there are two subsequences with very
different $p$, one is close to 1.1, the other is close to 2.0
according to \citet{kisslinger199107}.).  If the sequence is
sufficiently long, we truncate it and keep the part before the
distortion for the analysis.  Otherwise, we discard it.

3. The number of events is too small (say $<15$) or the duration is
too short (say $<1$ day), for example, events 16 and 18 last 0.21
days, events 13 and 39 last 15 days.

Of the 39 aftershock sequences studied by \citet{kisslinger199107}, 12
were thus excluded from our analysis. For the other 27 sequences we
first change the magnitude cutoff above which the aftershocks are
kept in the analysis.  This allows us to determine the best magnitude
cutoff for which the oscillations decorating Omori's power law for
the normalized cumulative number of events are the most regular. This
selection introduces a slight bias in favor of the presence of
log-periodicity, which in fact strengthens our negative conclusion
obtained by this approach. In any case, similar results are obtained
by choosing the magnitude threshold used by \citet{kisslinger199107}.
The regularity and strength of the oscillations are quantified by the
ratio between the sum of the squares of
the errors (RSSE) in the time domain resulting from the power law fit and
from a power law with log-periodic oscillations added on it.  The
presence of log-periodic oscillations is also quantified by the
significance level of the main peak of the oscillation spectrum
obtained from the Lomb method.

\begin{figure}
  \psfrag{t}[][]{$t$ (day)}
  \psfrag{Normalized CN}[][]{Normalized cumulative number}
  \figbox*{\hsize}{}{\includegraphics[width=8cm]{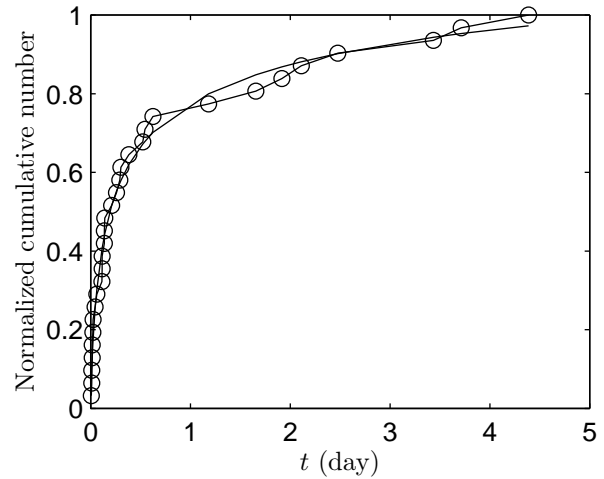}}
  \caption{Cumulative number of events at the optimal magnitude cutoff
    4.5 for the 1933 Long Beach earthquake aftershock sequence. The
    empirical data are represented by circles connected by lines
    to guide the eye.}
  \label{calaft1a.eps}
\end{figure}

\begin{figure}
  \psfrag{log\(t\)}[][]{$\ln(t)$}
  \psfrag{Oscillations}[][]{Oscillations}
  \figbox*{\hsize}{}{\includegraphics[width=8cm]{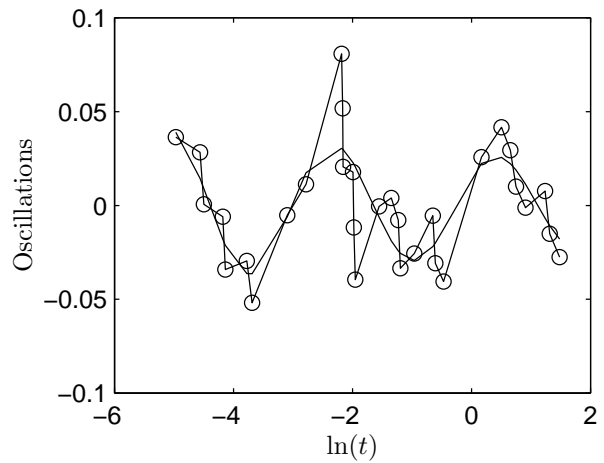}}
  \caption{Residual oscillations of the fit of Figure \ref{calaft1a.eps}
    by a simple power law are themselves fitted by a simple cosine
    function (solid line). The empirical data are represented by
     circles connected by lines to guide the eye.}
  \label{calaft1c.eps}
\end{figure}

\begin{figure}
  \psfrag{w}[][]{$\omega$}
  \psfrag{Lomb Power}[][]{Lomb power}
  \figbox*{\hsize}{}{\includegraphics[width=8cm]{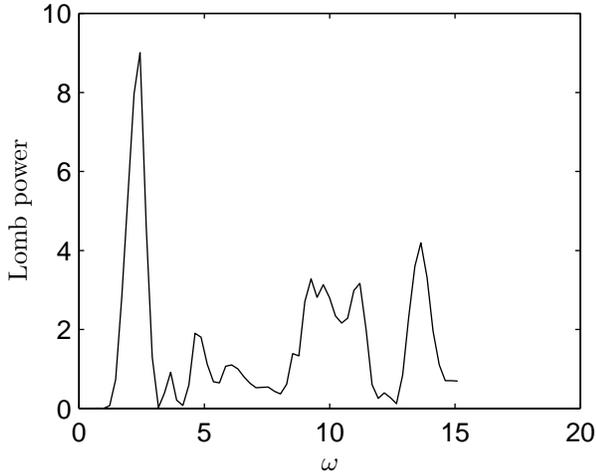}}
  \caption{Spectral analysis of the
    oscillations shown in Figure \ref{calaft1c.eps} using the Lomb
    Periodogram for unevenly sampled data.}
  \label{calaft1b.eps}
\end{figure}

We illustrate this procedure on the 1933 Long Beach earthquake
sequence, with a main shock magnitude of 6.3, which contains 571
events with minimum magnitude 2.0 and total duration of 2576 days.
For this sequence, using a magnitude cutoff of 4.5 provides the
strongest signal: we find the smallest correction for the origin of
time of the aftershock sequence (a fit to the modified Omori's law
$1/(t+C)^p$ gives a very small $C$, an exponent $p$ close to $1$ and a
high RSSE). The Lomb periodogram of the oscillations has also a very
significant frequency peak. The cumulative number of events is shown
in Figure \ref{calaft1a.eps}, in which the solid line is the fit
to a pure power law. The residual of this fit is shown in Figure
\ref{calaft1c.eps} and is fitted to a simple cosine function. The
corresponding spectral analysis is shown in Figure \ref{calaft1b.eps}.


\begin{figure}
  \psfrag{w}[][]{$\omega$}
  \psfrag{height of lomb peak}[][]{Height of Lomb peak}
  \figbox*{\hsize}{}{\includegraphics[angle=90,width=8cm]{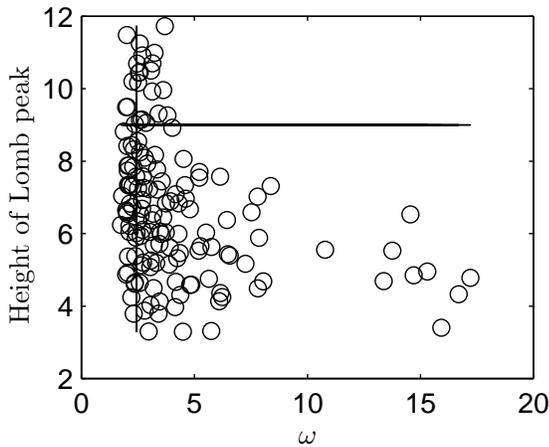}}
  \caption{Distribution of the log angular frequency $\omega$ and peak
    height of 200 synthetic data sets and comparison to the real
    sequence identified as the intersection of the two lines.}
  \label{calaft1shortsyn1mcb200alllmfq.eps}
\end{figure}

Synthetic tests were performed for this magnitude cutoff of 4.5 by
generating 200 numerical sequences with exactly the same parameters as
determined by the simple power law fit.  Figure
\ref{calaft1shortsyn1mcb200alllmfq.eps} shows that the real sequence
(defined as the intersection of the two lines) exhibits a log-periodic
frequency essentially equal to the most probable log frequency of the
synthetic data, with a Lomb peak not so uncommon either (corresponding
approximately to a 10\% significance level).

\begin{figure}
  \psfrag{Fluctuations around the Omori}[][]{\ \ \ \ \ \ \ \ Fluctuations around
    Omori's law}
  \psfrag{t}[][]{$t$}
  \figbox*{\hsize}{}{\includegraphics[width=8cm, bb=49 400 290 598,
clip]{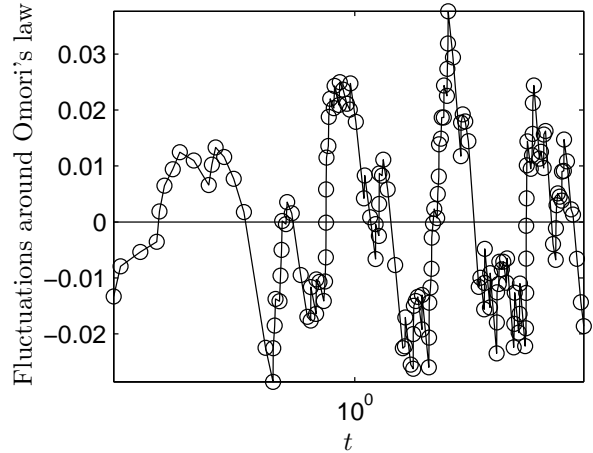}}
  \caption{Log-periodic oscillations around the cumulative number
    of events of the Santa Barbara Island aftershock sequence 30.}
  \label{x04271998_30.eps-a}
\end{figure}

\begin{figure}
  \psfrag{w}[][]{$\omega$}
  \psfrag{P}[][]{\ \ $P_N(\omega)$}
  \figbox*{\hsize}{}{\includegraphics[width=8cm, bb=316 400 547 598,
clip]{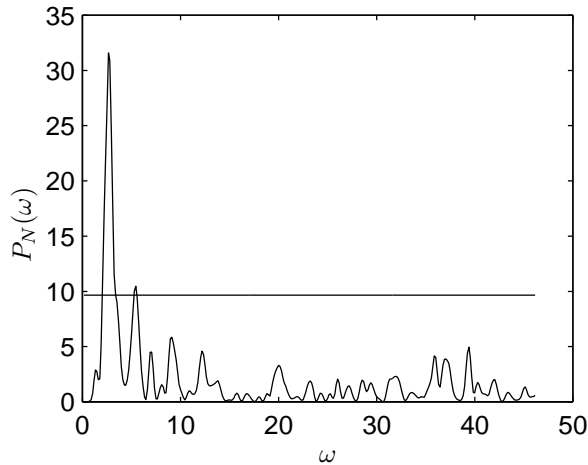}}
  \caption{Normalized Lomb periodogram of the oscillations in
      Figure \ref{x04271998_30.eps-a}. The horizontal line marks
      significance level of $0.01$.}
  \label{x04271998_30.eps-b}
\end{figure}

Aftershock sequence 30 of the Santa Barbara Island earthquake is
found to exhibit the strongest log-periodic oscillations, shown in
Figure \ref{x04271998_30.eps-a} and \ref{x04271998_30.eps-b}.
However, from 10 synthetic sequences using the parameters from the
real sequence, we observed four better log-periodic oscillations at
similar log frequencies. We also find that the amplitude of
oscillations of the real sequence is close to that of the synthetic
sequences, and this is a general property found for all the other
sequences.  From more extensive simulations with 200 synthetic
sequences we confirm that these oscillations shown in Figure
\ref{x04271998_30.eps-a} are near the most probable region of the
distribution of the log frequencies in the synthetic time series.

To summarize these results, let us first discuss the time series
obtained by using the magnitude cutoffs chosen by
\citet{kisslinger199107}.  We find that only 10 sequences among the 27
analyzed sequences have log frequencies outside the most probable
range of synthetic log frequencies.  Among them, seven (sequence 11,
13, 20, 22, 28, 29, 30) have log frequencies slightly higher than the
most probable synthetic frequencies and might thus be considered as
potential candidates for a genuine physical origin. The three other
cases have log frequencies that are lower than found for most of the
synthetic tests, a signature of a trend or large-scale aliasing.  With
respect to the statistical significance of the peaks obtained from the
Lomb periodogram, only six sequences (10, 18, 23, 27, 31, 36) have
spectral peaks higher than the most probable synthetic peak heights,
and none of them overlap with the seven candidates selected above.

For time series constructed with the best magnitude cutoff we find
that eight of the log frequencies are higher than those found at the
magnitude cutoffs of \citet{kisslinger199107}, nine are lower, and ten
are equal. However, the log frequencies from the corresponding
synthetic data sets are also shifted accordingly. Five sequences (5,
11, 15, 23, 30) have log frequencies higher than the most probable
range of the frequencies from the synthetic data sets, six sequences
(1, 13, 14, 22, 27, 36) have peak higher than those from synthetic
data sets, but there is no overlap between these two sets of five and
six sequences, respectively. No significant improvement was thus gained
by optimizing the magnitude cutoffs for more regular log-periodic
oscillations.  These results are summarized in Tables
\ref{omegatable1} and \ref{omegatable2}.  When our choice of the best
magnitude cutoff is the same as that of Kisslinger and Jones, the
sequence is excluded from Table \ref{omegatable2} to avoid repetition.

\begin{table*}
  \tablewidth{35pc}
  \caption{Log-Periodic Signals in the Oscillations Around the Modified
    Omori's Power Law of the Cumulative Number of Aftershocks Using
    Kisslinger and Jones' Magnitude Cutoff}
  \label{omegatable1}
  \begin{center}
    \begin{tabular}{ccccccccc}\\[-3ex]\hline\\[-1ex]
seq \#\tablenotemark{\it a}\tablenotetext{\it a}{The number of the data sets used by \citet{kisslinger199107}.}&
mm\tablenotemark{\it b}\tablenotetext{\it b}{Main shock magnitude.}&
$\omega_0$\tablenotemark{\it c}\tablenotetext{\it c}{The left-wise angular frequency at half the maximum of
the distribution of angular frequencies from synthetic aftershocks.}&
 $\omega_1$\tablenotemark{\it d}\tablenotetext{\it d}{The right-wise angular frequency at half the maximum of
the distribution of angular frequencies from synthetic aftershocks.}&
 $\omega$\tablenotemark{\it e}\tablenotetext{\it e}{From the real aftershock sequence.}&
 pkht$_0$\tablenotemark{\it f}\tablenotetext{\it f}{The left-wise angular frequency at half the maximum of the
 distribution of the peak heights of the most significant Lomb peak from
synthetic aftershocks.}&
 pkht$_1$\tablenotemark{\it g}\tablenotetext{\it g}{The right-wise angular frequency at half the maximum of
 the distribution of the peak heights of the most significant Lomb peak
 from synthetic aftershocks.}&
pkht\tablenotemark{\it h}\tablenotetext{\it h}{Peak height of the Lomb periodogram for the real aftershock
sequence.}&
Pr(Chance)\tablenotemark{\it i}\tablenotetext{\it i}{Probability of the log-periodicity in real aftershocks
being chance.}
      \\[1ex]\hline\\[-1ex]
01 &6.3&  1.0740  & 2.7077  &1.7175  &  14.4865 &  35.6532  & 14   &0.63
\\
02 &5.1&  1.4913  & 4.2834  &2.1367  &   1.6963 &   6.9105  & 5    &0.1974
\\
03 &6  &  1.8340  & 2.5803  &2.4112  &   4.0659 &   7.8017  & 5.5  &0.4300
\\
05 &6.9&  2.2869  & 2.8769  &1.8280  &   3.7103 &   8.1253  & 8    &0.1548
\\
06 &5.9&  0.7297  & 1.7442  &0.8516  &   8.0163 &  17.4935  & 9    &0.8495
\\
09 &6.2&  1.1510  & 1.4220  &1.2025  &  22.4059 &  29.2246  & 28   &0.3152
\\
10 &5.5&  1.5570  & 2.6391  &1.6537  &   9.7964 &  22.9913  & 37   & 0.005
\\
11 &6  &  0.8838  & 1.8858  &2.3547  &  13.0106 &  25.0258  & 16   &0.0606
\\
12 &5.9&  1.2057  & 2.9265  &1.5851  &  11.9195 &  26.1618  & 18   &0.6
\\
13 &5.7&  1.2308  & 1.7628  &2.1031  &   3.4584 &   7.6336  & 5.8  &0.2479
\\
14 &7.7&  1.1584  & 1.3052  &1.1763  &  31.2361 &  58.8183  & 44   &0.4724
\\
15 &6.2&  0.8841  & 2.1099  &2.0406  &   9.4233 &  22.2794  & 16   &0.0414
\\
17 &5.6&  1.6172  & 2.4806  &0.9766  &   1.4954 &   3.7202  & 3.7  &0.28
\\
18 &6.5&  0.9946  & 2.3998  &0.9688  &  11.3033 &  18.0648  & 26   &0.0118
\\
20 &6.6&  0.6763  & 1.6025  &2.0684  &  19.4909 &  46.3860  & 25   &0.0863
\\
22 &5.1&  1.1030  & 2.2221  &3.6476  &  38.4192 &  47.8631  & 37   &0.0125
\\
23 &5.8&  1.1490  & 2.6941  &1.4775  &   9.1692 &  17.2860  & 18   &0.1907
\\
24 &5  &  1.1602  & 2.2417  &1.2623  &  13.9662 &  28.9842  & 17   &0.7714
\\
25 &5.2&  1.6197  & 2.9343  &1.9517  &  31.0554 &  50.2901  & 42   &0.3984
\\
27 &5.5&  1.0248  & 2.3511  &1.1171  &   4.7256 &  15.4207  & 16   &0.0914
\\
28 &6.1&  1.9769  & 5.0416  &6.2079  &   2.6342 &   9.3955  & 8    &0.0102
\\
29 &5.7&  2.1068  & 3.3735  &3.4648  &   6.8699 &  18.2430  & 14   &0.0278
\\
30 &5.3&  1.1045  & 2.2538  &2.6459  &  22.4250 &  45.6508  & 32   &0.0408
\\
31 &6.5&  0.7557  & 1.8494  &1.1278  &  39.3589 &  68.9044  & 75   &0.0811
\\
32 &5.8&  1.1349  & 2.5303  &1.8404  &  31.0060 &  53.6768  & 36   &0.3254
\\
33 &5.9&  1.2886  & 2.3381  &1.7686  & 157.7374 & 430.4343  & 400  &0.0950
\\
36 &6  &  1.0569  & 2.0341  &1.4477  &  25.5461 &  51.6242  & 62   &0.0308
      \\[1ex] \hline\\[-3ex]
    \end{tabular}
  \end{center}
\end{table*}
\begin{table*}
  \tablewidth{35pc}
  \caption{Log-Periodic Signals in the Oscillations Around the Modified
    Omori's Power Law of the Cumulative Number of Aftershocks Using Other
    Magnitude Cutoff\tablenotemark{\it a}\tablenotetext{\it
    a}{Meanings as in Table 
    \ref{omegatable1}.}}
  \label{omegatable2}
  \begin{center}
    \begin{tabular}{cccccccc}\\[-3ex]\hline\\[-1ex]
seq \#&
$\omega_0$&
 $\omega_1$&
 $\omega$&
 pkht$_0$&
 pkht$_1$&
pkht&
Pr(Chance)
      \\[1ex]\hline\\[-1ex]
01 &  1.7518  &  3.6436 & 2.4358   & 3.3245  &   8.6912   &  9  &0.1181
\\
03 &  1.0438  &  3.3507 & 1.9949   & 6.0207  &  12.8901   &  11 &0.0417
\\
05 &  1.0686  &  2.8314 & 11.9837  & 4.5386  &  10.1490   &  8  &0.0068
\\
11 &  1.0552  &  2.3074 & 2.3547   & 9.5031  &  17.9494   &  16 &0.0606 
\\
13 &  1.2473  &  2.8182 & 2.1031   & 3.3596  &  10.3787   &  11 &0.0213
\\
14 &  1.0120  &  2.1585 & 1.04     &10.1416  &  22.0273   &  28 &0.0357
\\
15 &  0.9610  &  2.0448 & 2.2577   &12.5354  &  34.1214   &  30 &0.0132
\\
17 &  1.1423  &  3.4385 & 1.0949   & 1.3481  &   6.9516   &  6  &0.375 
\\
18 &  1.0009  &  2.5511 & 2.5486   &15.9492  &  34.7236   &  30 &0.0301
\\
20 &  0.7923  &  1.8448 & 1.3063   &52.6478  & 118.4051   &  95 &0.1019
\\
22 &  1.7097  &  3.0411 & 1.0599   &50.9766  &  71.1400   &  95 &0.005 
\\
23 &  1.4364  &  2.4938 & 2.6761   &19.0478  &  56.8038   &  40 &0.0455
\\
24 &  1.3325  &  2.6773 & 1.2826   & 7.5998  &  17.1631   &  15 &0.3005
\\
25 &  1.6782  &  3.6164 & 1.7050   &49.2585  &  71.3087   &  10 &0.005 
\\
27 &  0.9257  &  2.2607 & 1.0769   & 7.9868  &  22.5240   &  24 &0.0655
\\
28 &  1.7700  &  3.8280 & 2.4337   &42.6775  &  78.4212   &  12 &0.005 
\\
29 &  1.8001  &  3.1963 & 2.4748   &12.6694  &  26.2655   &  25 &0.1224
\\
30 &  0.9557  &  2.4180 & 2.6459   &16.8089  &  33.8548   &  31 &0.0408
\\
32 &  1.1580  &  2.7235 & 1.6682   &13.4569  &  33.1647   &  22 &0.3408
\\
33 &  0.9972  &  2.8735 & 2.2487   & 3.4352  &   9.3050   &  9  &0.0588
      \\[1ex] \hline\\[-3ex]
    \end{tabular}
  \end{center}
\end{table*}

We thus find that the 27 aftershock sequences in southern California
analyzed by the cumulative distribution method exhibit clear
log-periodic structures, which are, however, not strongly statistically
significantly different from those generated in synthetic time series.
Indeed, synthetic series generated using the pure Omori's power law
exhibit similar structures that result from the noise reddening
mechanism explained in sections \ref{noise} and \ref{ns}.
Optimization over magnitude cutoff for more regular oscillations does
not significantly improve the results. We observed that log-periodic
oscillations with angular log frequencies are found in the interval
$[0.8, 6.2]$ in most sequences with various regularity.  When compared
with synthetic aftershock sequences, these frequencies and the power
spectrum at these frequencies are found among the most probable values
in synthetic sequences with the same parameters and analyzed in the
same way. The amplitudes of the oscillations in the real sequences are
almost equal to those of the synthetic data. 

\begin{figure}
  \psfrag{Log-derivative}[][]{log derivative}
  \psfrag{log\(t\)}[ct][cB]{$\ln(t)$}
  \figbox*{\hsize}{}{\includegraphics[width=8cm]{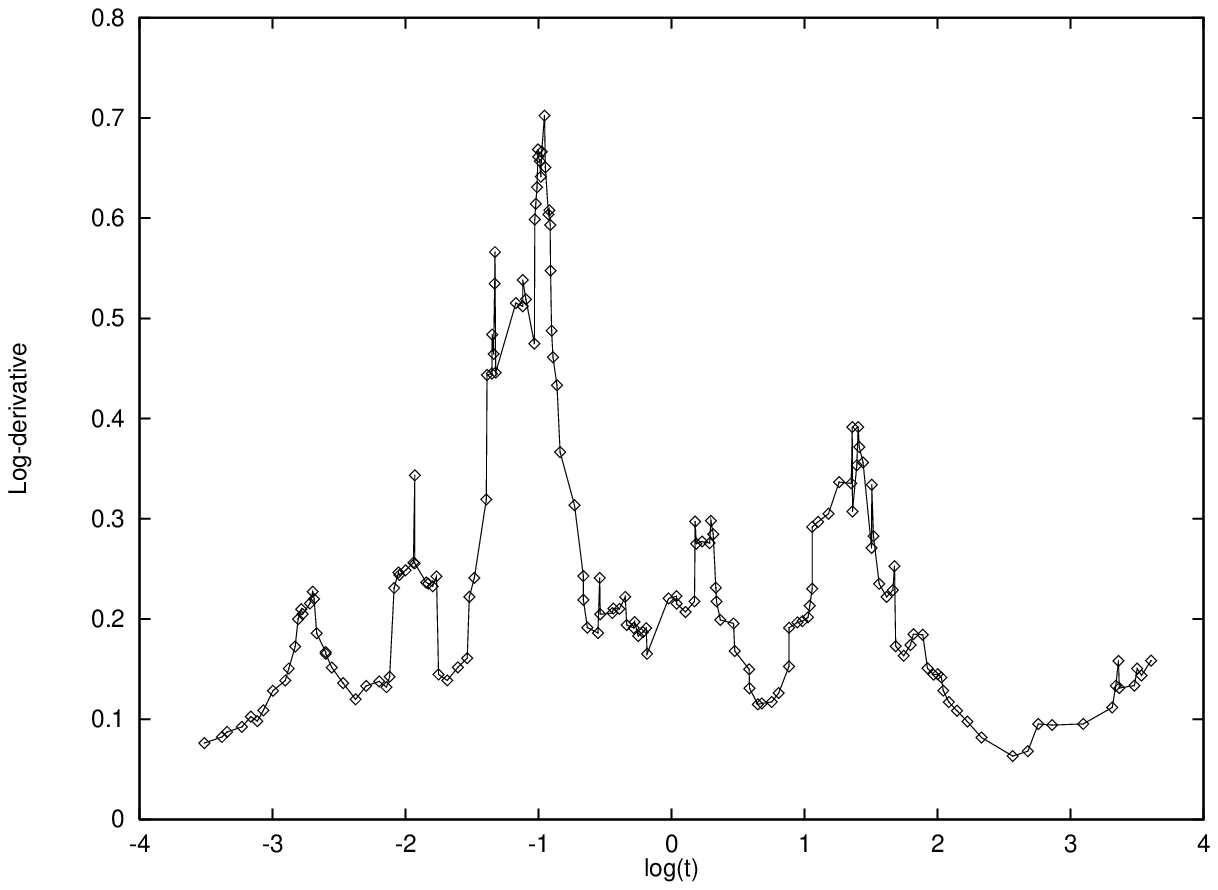}}
  \caption{Aftershock sequence following the Owens
    Valley $M5.8$ earthquake on October 14, 1978. 
    Derivative (denoted ``log derivative'') of the cumulative number
    of aftershocks with respect to the logarithm of time as a function
    of $\ln\lp t\rp$ is shown.}
  \label{realas}
\end{figure}

\begin{figure}
  \psfrag{Log-derivative}[][]{Log derivative}
  \psfrag{log\(t\)}[ct][cB]{$\ln(t)$}
  \figbox*{\hsize}{}{\includegraphics[width=8cm]{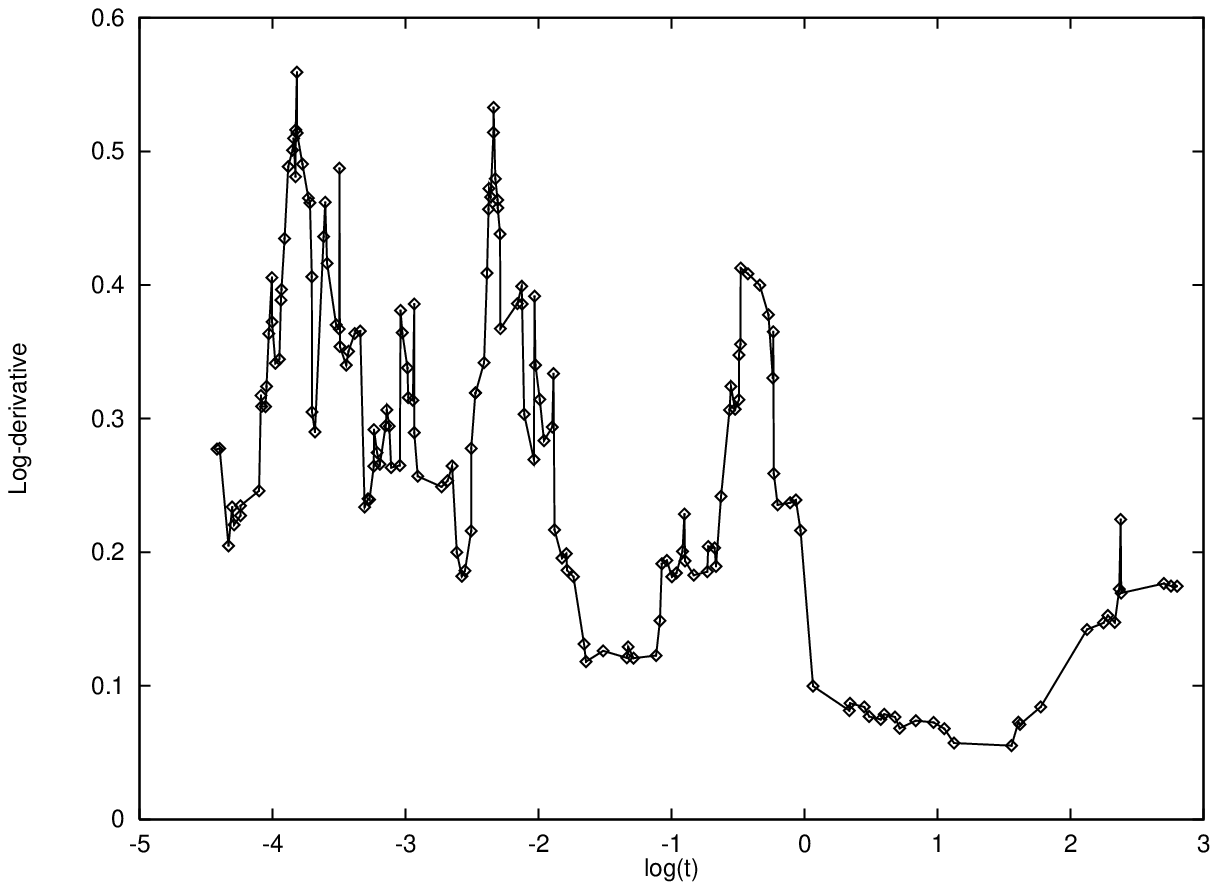}}
  \caption{Same as Figure \ref{realas}
    for a synthetic data set corresponding to the aftershock sequence
    shown in Figure \protect\ref{realas}. The same number of points
    and time interval as in Figure \protect\ref{realas} were used. The
    $p$ value was $p\approx 1.249$ as listed by
    \citet{kisslinger199107}.}
  \label{syntas}
\end{figure}

Another natural quantity to investigate in order to qualify Omori's
law (\ref{omorilaw}) is to take the numerical derivative of the
cumulative number ${\cal N}$ of aftershocks with respect to the
variable $\ln t$. For a pure Omori's law with exponent $p=1$, this
log derivative should be constant.  Any deviations from a constant
provides a measure of deviation from Omori's law. We estimate the
logarithmic derivative by
\be
\frac{d{\cal N}}{d\ln\lp t\rp}\Bigg|_{t=t_i} =
\frac{{\cal N}\lp t_{i+1}\rp  - \ln {\cal N} \lp t_{i-1}\rp }
{\ln\lp t_{i+1} \rp - \ln \lp t_{i-1}\rp }~.
\ee
Owing to the intermittent nature of the data, a moving average has been
applied to the derivative. An example of the result is shown in Figure
\ref{realas}. Rather than a constant that would qualify Omori's law,
we observe quite strong approximately log-periodic signatures.
However, a test on a synthetic data sequence, using the same number of
points and time window as well as the same Omori's $p$ estimated by
\citet{kisslinger199107}, gives a similar signal, as shown in
Figure \ref{syntas}, namely approximately equidistant peaks of
fluctuations. The log-periodic signatures are generated from the noise
and are probably not genuine. In fact, the most obvious difference
between the true data and the synthetic data is that the oscillations in
the synthetic data have been shifted down to shorter timescales. This
is presumably due to an imprecise estimation of the value of the
exponent $p$.

\subsection{Cumulating and Log-Periodicity}
\label{destroy}

The essential property of the logarithmic sampling is a flat noise
spectrum (in the log variable) before integration. Integration then
leads to the appearance of a low-frequency peak, as discussed in
Section \ref{cumulsect}.
Different samplings will give rise to different noise spectra, which,
when convoluted with the integration, might well not create spurious
log-periodicity, or even destroy an existing one. We now give an
example of the latter situation.

We have generated $10$ data sets with $30$ points each using 
\be \label{surreq1}
y\lp t\rp = t^{-0.5}\left[ 1+ 0.1\cos\lp 7\ln t \rp \right]\,,
\ee
where the sampling was random in the sense that the spacing between
two consecutive points was chosen from the interval $\left[
  0,20\right]$ with uniform probability. From the Lomb periodogram of
the log derivative $[\ln y(t_{i+1}) - \ln y(t_{i})]/[\ln t_{i+1} - \ln
t_{i}]$ shown in Figure \ref{01f2}, we can clearly identify the
log-periodic component in the signal and its log frequency $f = {7
  / 2 \pi} \approx 1.1$ with a level of significance better than
$0.995$ (see Figure \ref{01f5}).

In contrast, the log-periodic signal completely disappears in the
cumulative distribution as seen in Figure \ref{01f6}.  This not only
shows that a strong log-periodic signal will be destroyed by
``integration'' but also that without noise, only two to three
oscillations are needed to qualify log-periodicity.

It is well-known that an integration of sampled data corresponds to a
low-pass filter. What we have seen here is that this low-pass
filtering extends into the log frequency domain with surprising
efficiency.

\begin{figure}
  \figbox*{\hsize}{}{\includegraphics[width=8cm]{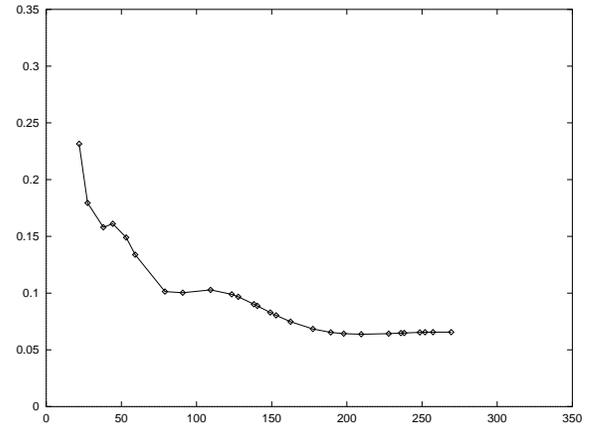}}
  \caption{Typical artificial data set of 30 points
    generated with Eq.~(\ref{surreq1}) with a random sampling
    such that the spacing between two consecutive points was chosen
    from the interval $[0,20]$ with a uniform probability.}
  \label{01f1}
\end{figure}
\begin{figure}
  \figbox*{\hsize}{}{\includegraphics[width=8cm]{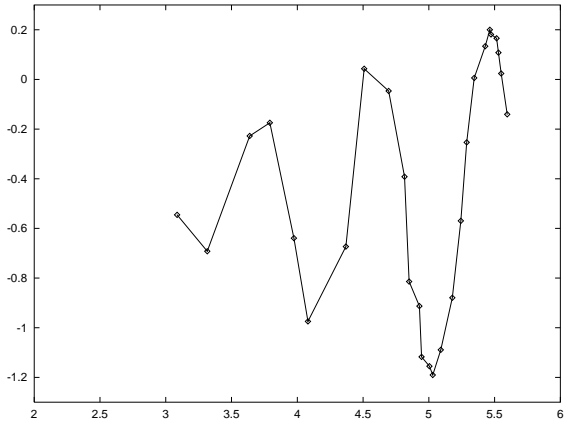}}
  \caption{The logarithmic derivative of the data set in
    Figure \protect\ref{01f1}.}
  \label{01f2}
\end{figure}
\begin{figure}
  \figbox*{\hsize}{}{\includegraphics[width=8cm]{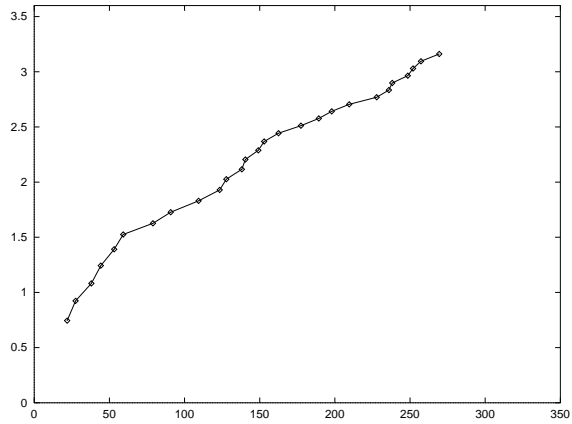}}
  \caption{The cumulative distributions of the data set in
    Figure \ref{01f1}.}
  \label{01f3}
\end{figure}
\begin{figure}
  \figbox*{\hsize}{}{\includegraphics[width=8cm]{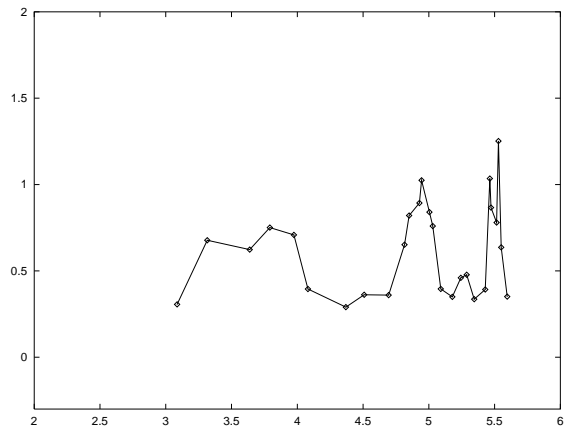}}
  \caption{The logarithmic derivative of the cumulative
    distributions in Figure \ref{01f3}.}
  \label{01f4}
\end{figure}
\begin{figure}
  \figbox*{\hsize}{}{\includegraphics[width=8cm]{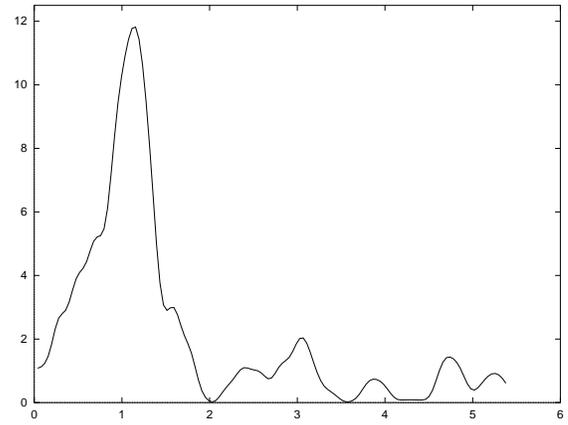}}
  \caption{The Lomb periodogram of the derivative in
    Figure \ref{01f2} at $f = {7 / 2 \pi} \approx 1.1$.}
  \label{01f5}
\end{figure}
\begin{figure}
  \figbox*{\hsize}{}{\includegraphics[width=8cm]{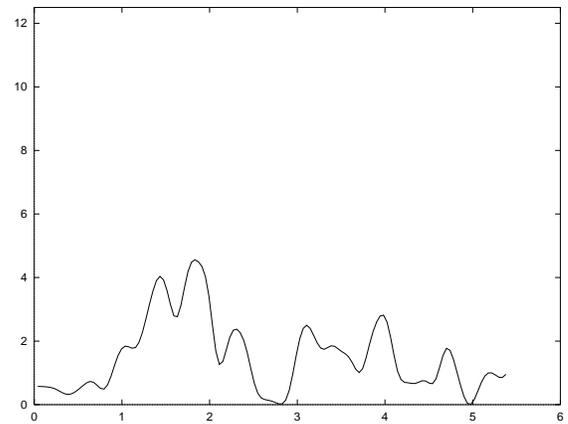}}
  \caption{The Lomb periodogram of the derivative in Figure
    \ref{01f4}.}
  \label{01f6}
\end{figure}

\section{Maximum Likelihood Analysis of the Aftershock Sequences}
\label{logbinning}

The previous analyses illustrate the danger of apparently innocuous
manipulations of the data, such as calculating cumulative
distributions of aftershock sequences.  This suggests that a method
that gives direct access to the local Omori's exponent may be better
suited for the unbiased detection of log-periodic structures. Indeed,
the most direct quantification of departure from Omori's law is to
look at the logarithmic derivative of the total number $N$ of events
until time $t$:
\bea \label{eqderiv}
\frac{d \ln N\lp t\rp}{d\ln t} = - p\lp t \rp ~,
\eea
which defines the local (\emph{i.e.}, at time $t$) Omori's exponent $p(t)$.

The maximum likelihood method is nonparametric and gives the most
probable value of $p$ given the data. In practice, we perform the
maximum likelihood determination in a moving window from $t$ to $t_U$.
Moving $t$ and $t_U$ over the time axis allows us to capture local
fluctuations in the Omori's $p$ and provides us with a method whereby
we can directly estimate any log-periodic oscillations \citep{leephd}.

\subsection{Maximum Likelihood Estimation of Omori's Exponent}

The maximum likelihood method estimates the $p$ value in the window
from $t$ to $t_U$ by maximizing the probability that the particular
temporal distribution of the $N$ events in that window results from
Omori's law with this particular $p$ value. The most likely $p$ value
is then given by the implicit equation
\be
{1 \over p-1} + {t_U^{p-1} \ln t - t^{p-1} \ln t_U \over t_U^{p-1} - t^{p-1}} =
 \langle \ln t \rangle_N ~,  \label{hajakk}
\ee
where the $t_n$ are the time occurrences of the $N$ events between $t$
and $t_U$ and
\be
\langle \ln t \rangle_N \equiv {1 \over N} \sum_{n=1}^N \ln t_n ~.
\ee
An expansion in powers of $(p-1) \ln{t_U / t}$ gives the explicit
solution
\be
p \approx 12\frac{\ln\sqrt{t t_U}- \langle \ln t \rangle_N }
{\lp \ln \frac{t_U}{t}\rp^2 +1 }~, \label{yajiiexactal}
\ee
with a typical error
\be
\sigma \approx \sqrt{\frac{12}{\left( N -1\right)
\left( \ln \frac{t_{U}}{t}\right)^{2}} }~,
\ee
obtained from the negative Hessian of the log likelihood. Only the
exponent $p$ is estimated from the formulas.  Expression
(\ref{yajiiexactal}) gives an extremely good approximation to the
exact solution of Eq.~(\ref{hajakk}), since the first nonzero
correction is proportional to the fourth power $\approx 10^{-2}~
[(p-1) \ln{t_U / t}]^4$ of the expansion parameter. We have
verified that the values $p$ given by (\ref{yajiiexactal}) are
indistinguishable from the exact solution of (\ref{hajakk}) for the
range of $p$ values of interest for aftershocks.  Since the error
$\sigma$ is inversely proportional to the inverse of the square of the
number $N$ of events in the time window, probing smaller timescales
is bought at a price of larger errors.

\subsection{Generation of Log-Periodicity}

\begin{figure}
  \psfrag{Time }[ct][cB]{Time }
  \psfrag{p-value}[][]{$p$ value}
  \figbox*{\hsize}{}{\includegraphics[width=8cm]{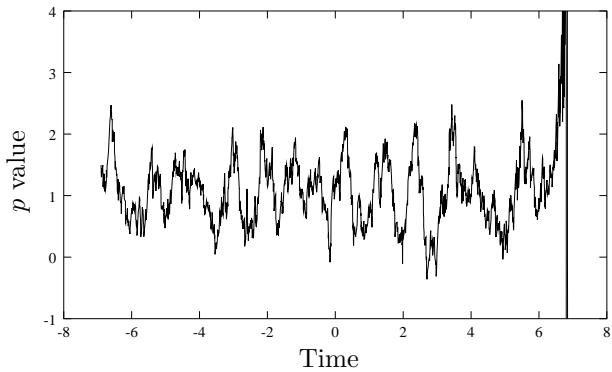}}
  \caption{Omori's exponent $p$
    as a function of the logarithm of time in a running window of
    $N=100$ events for a synthetic aftershock time series generated
    using (\ref{hgahaka}) with $p=1.0$, extending from time $1$ to
    $1000$ and containing  $\sim 500$ events.  Log-periodicity
    is clearly evident to the naked eye.}
  \label{pval-p=3D1}
\end{figure}

\begin{figure}
  \psfrag{Power}[][]{Lomb power}
  \psfrag{Frequency}[ct][cB]{Frequency}
  \psfrag{synthetic with p=1.0, 2000 events}[][]{}
  \figbox*{\hsize}{}{\includegraphics[width=8cm]{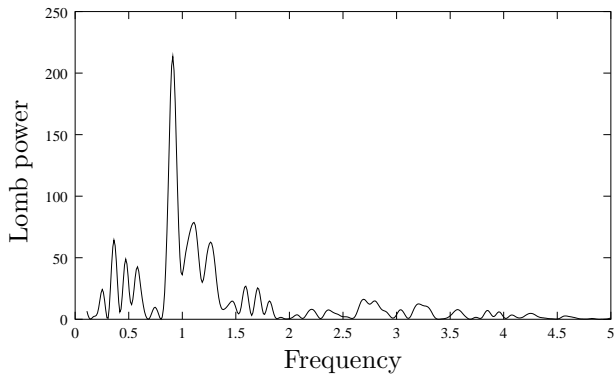}}
  \caption{Lomb periodogram of Omori's
    exponent $p$ shown in Figure \ref{pval-p=3D1} exhibiting a highly
    significant peak.}
  \label{pval-p=3D1b}
\end{figure}

Since the precision on the determination of $p$ is essentially
controlled by the number $N$ of points in the running window, it seems
natural to vary $t_U$ as a function of $t$ such that the number $N$ of
points in the running window remains fixed. Any other specification
deteriorates the homogeneity of the determination of the exponent
along the time axis.  However, this procedure turns out to generate
strong spurious log-periodicity as well.  This is illustrated in
Figure \ref{pval-p=3D1} showing the local $p$ value in a running
window with $N=100$ events determined from a synthetic aftershock
series generated using (\ref{hgahaka}). The spectral analysis of this
data set shown in Figure \ref{pval-p=3D1b}, confirms the statistical
significance of the log-periodic signal.

By construction, the synthetic aftershock sequence obeys exactly
Omori's law with $p=1$, and any observed structure decorating this
simple power law reflects random noise and finite size effects. We
trace back the log-periodicity observed in Figure \ref{pval-p=3D1}
to the constraint that the number of points $N$ in the analyzing
window is fixed. Indeed, for Omori's law with $p=1$, this corresponds
exactly to a sampling that is uniform in log time. As for the
cumulative method, the most probable component of the noise has a
significant spectral power.

\subsubsection{Log frequency as a function of window size.}

\begin{figure}
  \psfrag{Normalized Power}[][]{Normalized Lomb power}
  \psfrag{Frequency}[ct][cB]{Frequency}
  \figbox*{\hsize}{}{\includegraphics[width=8cm]{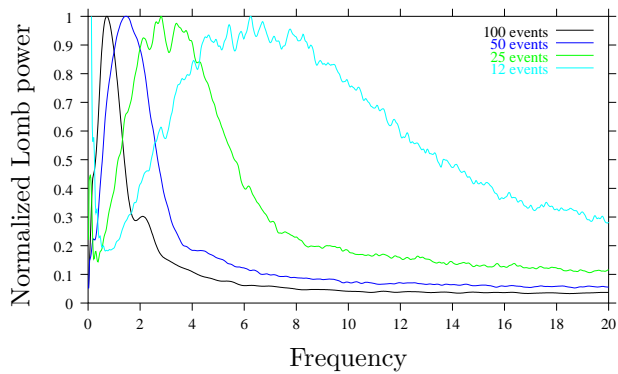}}
  \caption{Average Lomb periodogram  over $1000$
    synthetic aftershock realizations with $570$ events and $p=1.282$
    of the function $p(t)$ calculated with the maximum likelihood
    method in running windows of size $N=12,
    25, 50,$ and $100$ events, respectively.}
  \label{windowkjalo}
\end{figure}

\begin{figure}
  \psfrag{Normalized Power}[][]{Normalized Lomb power}
  \psfrag{Frequency}[ct][cB]{Frequency}
  \figbox*{\hsize}{}{\includegraphics[width=8cm]{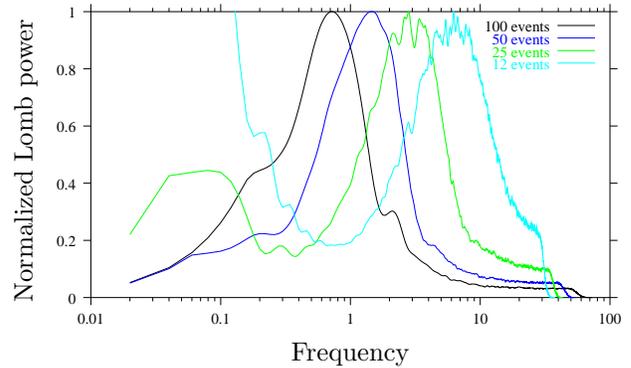}}
  \caption{Same as Figure \ref{windowkjalo} using
    a log scale for the log frequencies illuminating the very similar
    approximate lognormal structure of the Lomb periodograms.}
  \label{windowkjaaqalo}
\end{figure}

To demonstrate the origin of log-periodicity in the maximum likelihood
method with a fixed number of points in the running window, we
generate 1000 synthetic aftershock series with $570$ events and
$p=1.282$, corresponding to the values observed for the aftershock
sequence of Long Beach 1933 documented by \citet{kisslinger199107}.
For each of the 1000 synthetic sequences we have determined $p(t)$ in
running windows of size $N=12, 25, 50,$ and $100$ events.
For a fixed window size $N$ we took the Lomb periodogram of each
$p(t)$ and then averaged them over the 1000 realizations. We thus
obtain averaged Lomb periodograms for the four different values $N=12,
25, 50,$ and $100$, shown in the Figures \ref{windowkjalo} and
\ref{windowkjaaqalo}.  It is clear that the dominating log frequency
is a function of window size $N$.

\begin{figure}
  \psfrag{Frequency}[][]{Frequency}
  \psfrag{Window Size \(events\)}[][]{Window Size (events)}
  \figbox*{\hsize}{}{\includegraphics[width=8cm]{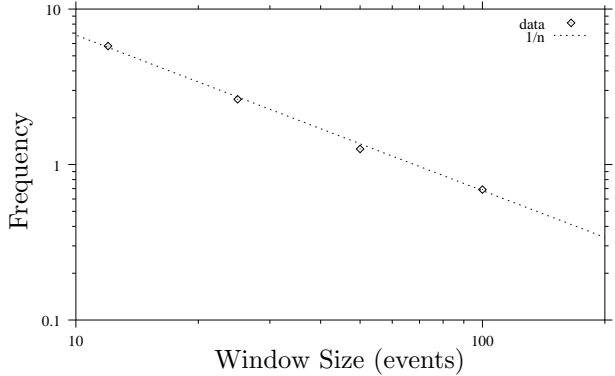}}
  \caption{Dependence of the most probable
    log frequency (in logarithmic scale) corresponding to the maximum
    of the peaks in Figures \ref{windowkjalo} and \ref{windowkjaaqalo}
    as a function of the logarithm of the window size $N$. The dashed
    line has a slope $-1$.}
  \label{windowkjalaao}
\end{figure}

The approximate lognormal shape of the Lomb periodograms shown in
Figure \ref{windowkjaaqalo} is remarkably consistent across the
different values of the window size and is characterized by
approximately the same standard deviation found close to $0.45$ for
all cases.  A plot of the logarithm of the log frequency versus the
logarithm of the window size $N$ is given in Figure
\ref{windowkjalaao} showing that the log frequency is inversely
proportional to $N$. For a point process with Omori's exponent close
to $1$, keeping a constant window size $N$ is approximately equivalent
to having windows with a constant value of $t_{U}/t$
where \( t_{U} \) and \( t \) are the upper and lower times of the
window:
\begin{eqnarray} \label{jajkaq}
f & \sim  & \frac{1}{N}\\
 & \simeq  & \left( \ln \frac{t_{U}}{t}\right) ^{-1}~,
\end{eqnarray}
where $N$ is window size.  This reveals that using a fixed number of
observations in each window corresponds to log sampling, a mechanism
already identified as a source of log-periodicity.

\begin{figure}
  \psfrag{Frequency}[][]{Frequency}
  \psfrag{Window}[ct][cB]{Window Size (events)}
  \figbox*{\hsize}{}{\includegraphics[width=8cm]{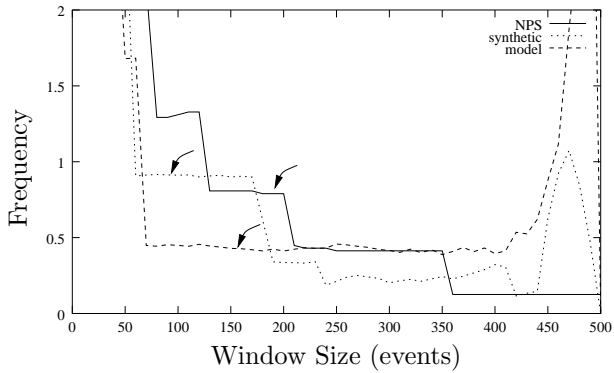}}
  \caption{Log frequency as a function of window size $N$
    for a synthetic sequence, a real sequence (North Palm Springs of
    1986 which contains 1437 events), and a sequence from the stress
    corrosion model studied by \citet{lee9903}.}
  \label{sweep}
\end{figure}

Of course, for a single realization, fluctuations scramble the
clear-cut result (\ref{jajkaq}). This is illustrated in Figure
\ref{sweep} which tests the dependence of the most probable
log frequencies of the $p(t)$ function as a function of window size
$N$ varying from $10$ to $500$.  Figure \ref{sweep} plots the most
probable log frequency as a function of window size $N$ for three
cases. The first one corresponds to a synthetic sequence of $2000$
events with exponent \( p=1.0. \) The second one is for a real
aftershock sequence, the North Palm Springs sequence of 1986, which
contains 1437 events with $p\approx 1.14$.  The last example is a
numerical simulation of $2000$ events from the $p=1$ stress corrosion
model of aftershocks studied \citep{lee9903} with conservation
parameter $\gamma =0.15$, stress corrosion exponent $\alpha =15$, and
stress drop parameter $\beta =0.999$, for which log-periodicity
decorating Omori's law is very strong and thus strikingly visible
directly on the $p(t)$ function (not shown).  At large window sizes,
multiple periods of log oscillation are included within a window, and
Omori's exponent $p$ starts to average out giving a very low value
for the log frequency.  At very small windows, small-scale variations
dominate yielding anomalously high log frequencies. The optimal window
sizes are small enough to cover less than a whole period of
oscillation in order to resolve the oscillations but large enough not
to be dominated by the small-scale variations.  For the numerical
simulation of the stress corrosion model of aftershocks studied by
\citet{lee9903}, we observe a well-defined plateau for the
log frequency as a function of window size $N$, extending from $N=60$
to above $400$, qualifying a genuine log-periodicity\,: the
theoretical value of the log frequency is \( 0.41, \) \citep{lee9903},
which is consistent with the measured values.  In contrast, the
log frequency dependences of the synthetic Omori's law and of the
North Palm Springs sequence are very similar to each other and show an
overall decay as predicted by (\ref{jajkaq}), decorated by
fluctuations.

\subsubsection{Log frequencies as a function of Omori's exponent $p$.}

\begin{figure}
  \psfrag{p-value}[][]{$p$ value}
  \psfrag{Time}[ct][cB]{Time}
  \psfrag{synthetic with p=1.5, 2000 events}[][]{}
  \figbox*{\hsize}{}{\includegraphics[width=8cm]{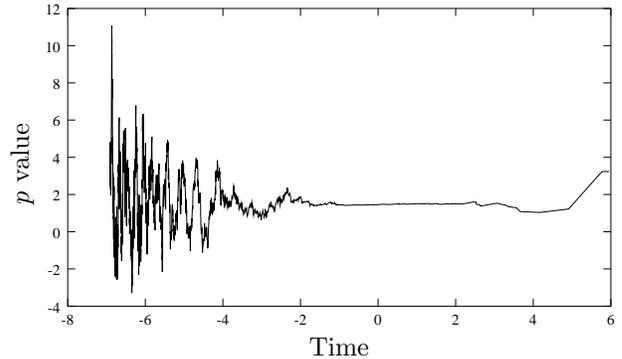}}
  \caption{$p(t)$ as a function of
    the logarithm of time for a synthetic sequence generated with
    $p=1.5$, 2000 events.}
  \label{pval-1.5and0.5a}
\end{figure}

\begin{figure}
  \psfrag{p-value}[][]{$p$ value}
  \psfrag{Time}[ct][cB]{Time}
  \psfrag{synthetic with p=0.5, 2000 events}[][]{}
  \figbox*{\hsize}{}{\includegraphics[width=8cm]{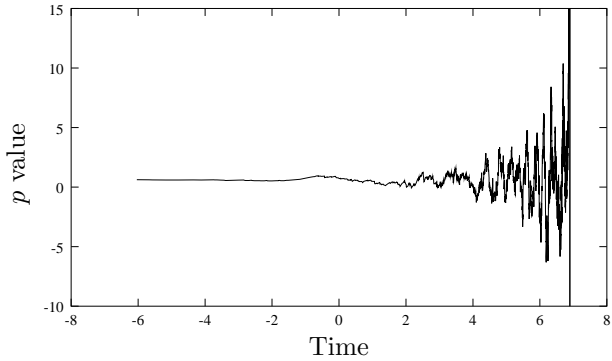}}
  \caption{$p(t)$ as a function of
    the logarithm of time for a synthetic sequence generated with
    $p=0.5$, 2000 events.}
  \label{pval-1.5and0.5b}
\end{figure}

Varying the value of Omori's exponent $p$ away from $1$ yields
quantitatively different results. The functions $p(t)$ measured from
synthetic time series generated with $p=1.5$ and $p=0.5$ are given in
Figures \ref{pval-1.5and0.5a} and \ref{pval-1.5and0.5b}.
Three features can be observed: (1) the log frequency has increased in
both cases; (2) the oscillations are no longer truly log-periodic
since the log frequency decreases with time for $p=1.5$ and increases
with time for $p=0.5$; and (3) the amplitude of oscillations changes
with time in both cases. To recover regular oscillations, we see from
(\ref{hgahaka}) that one would need to plot $p(t)$ as a function of
$t^{1-p}$ instead of $\ln t$, which is the correct variable only for
series with constant $p=1$.


\begin{figure}
  \psfrag{Frequency}[][]{Frequency}
  \psfrag{p-value}[ct][cB]{$p$ value}
  \figbox*{\hsize}{}{\includegraphics[width=8cm]{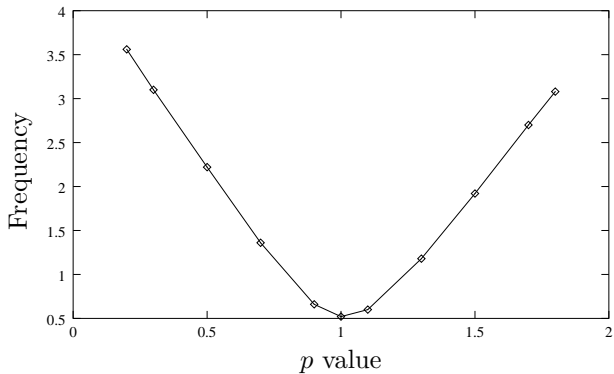}}
  \caption{Average log frequency as a function of $p$ value for synthetic
    aftershock sequences.}
  \label{freq-vs-p-synthetic}
\end{figure}

Notwithstanding its variation with time, we determine the log
frequency of the dominant peak by averaging over the Lomb periodograms
of $100$ realizations of synthetic data. The dependence of the average
log frequency as a function of the exponent $p$ used in generating
synthetic aftershock sequences is presented in Figure
\ref{freq-vs-p-synthetic}. For each $p$ value the average frequency
has been averaged over 100 realizations.  As $p$ differs from $1$, the
sequence exhibits higher and higher log frequencies.  Figure
\ref{freq-vs-p-synthetic} recovers the results already shown in Figure
\ref{wm} for the cumulative method.

\begin{figure}
  \psfrag{Frequency}[][]{Frequency}
  \psfrag{p-value}[ct][cB]{$p$ value}
  \figbox*{\hsize}{}{\includegraphics[width=8cm]{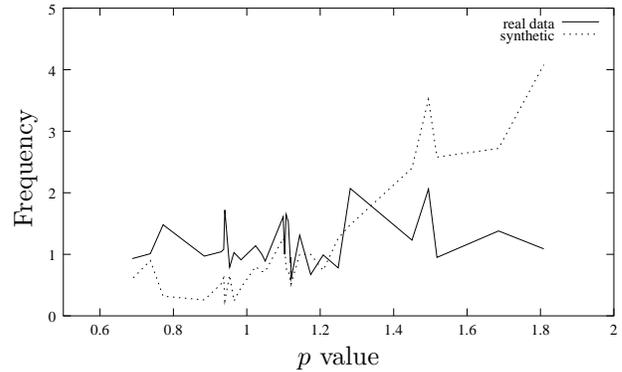}}
  \caption{Average log frequency as a function of Omori's exponent $p$
    in real and synthetic sequences. The confidence bounds are seen
    from the size of the fluctuations of the synthetic data.}
  \label{freq-vs-real}
\end{figure}

We have attempted to use the insight provided by Figures
\ref{windowkjalaao}, \ref{sweep}, and \ref{freq-vs-p-synthetic} to
distinguish the log-periodic signatures observed in real aftershock
sequences from those obtained in synthetic data sets with the maximum
likelihood approach.  We estimate the log frequency in the real
aftershock sequences as shown by the arrows in Figure \ref{sweep} on a
plateau in a central region for the window size which seems to provide
a reasonable compromise as explained above.  Each of the real
aftershock sequences is characterized by its average $p$ value
determined by \citet{kisslinger199107}. We use this $p$ value to
generate 1000 synthetic sequences with the same number of events. We
record the log frequency of the largest peak of the Lomb periodogram
of the function $p(t)$ for each of these 1000 synthetic sequences and
then average it. Spanning the different real aftershock sequences
allows us to span a relatively large interval of $p$ values. We thus
represent in Figure \ref{freq-vs-real} the most probable log frequency
as a function of average $p$ value obtained from the set of real
aftershock sequences and their corresponding synthetic ensemble.  As
in Figure \ref{freq-vs-p-synthetic}, the synthetic log frequency
average is increasing for $p >1$ (notwithstanding the averaging over
1000 realizations, fluctuations remain apparent). In contrast, the
most probable log frequency obtained for the real aftershock sequences
do not seem to be noticeably dependent on the average $p$. This
difference suggests that log-periodicity in real aftershock sequences
could be real.  However, the difference is weak and not clearly above
noise level. By itself, this result is suggestive but not enough to
substantiate the claim of log-periodicity in real data
\citep{lee9903}.

\subsection{Can Log-Periodicity Be ``Saved'' for Aftershocks?}

\subsubsection{Fixed time window size.}

\begin{figure}
  \psfrag{p-value}[][]{$p$ value}
  \psfrag{Time \(days\)}[ct][cB]{Time (days)}
  \figbox*{\hsize}{}{\includegraphics[width=8cm]{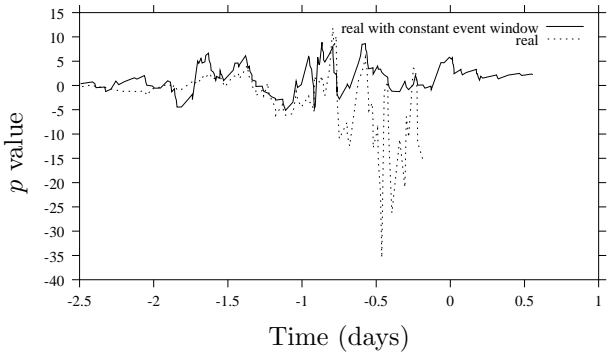}}
  \caption{Comparison of the local $p(t)$ as a
    function of the logarithm of time for event 29 (Westmoreland
    aftershock sequence) in the list of \citet{kisslinger199107} using
    the maximum likelihood method with a running window of fixed time
    duration of 0.1 day and with a fixed number of aftershocks
    (log sampling). We observe structures that are common to the two
    methods: this may be the signature of a real log-periodicity.}
  \label{event29fixedsin}
\end{figure}

\begin{figure}
  \psfrag{Frequency}[ct][cB]{Frequency}
  \psfrag{Normalized Power}[][]{Normalized Lomb power}
  \figbox*{\hsize}{}{\includegraphics[width=8cm]{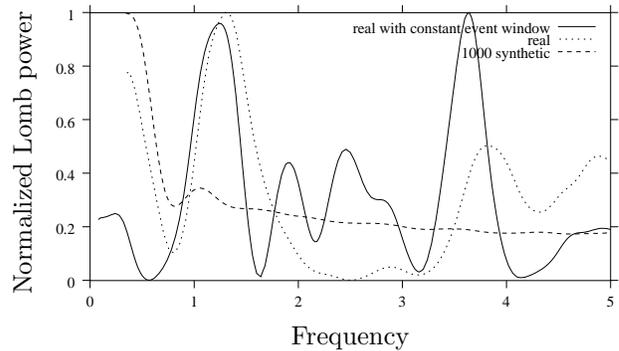}}
  \caption{Comparison of the Lomb
    periodograms corresponding to the two signals shown in Figure
    \ref{event29fixedsin}. We observe a common peak at a log frequency
    close to $1.2$.}
  \label{windowkjaasdgaaqalo}
\end{figure}

In view of the spurious log-periodicity generated when using a
constant number of points in a moving window which is equivalent to
log sampling, we redo all calculations using the maximum likelihood
approach with windows of fixed duration (for instance one-day
windows).  These calculations are performed on all sequences
documented by \citep{kisslinger199107} as well as on the more recent
Loma Prieta, Landers, and Northridge aftershock sequences.  As typical
examples, we show the 1981 Westmoreland and 1989 Loma Prieta
aftershock sequences. We choose the Westmoreland as an example with \(
p>1.4 \) and also with fluctuations greater than one standard
deviation ($2 \sigma$) from the mean equal to one (Figure
\ref{event29fixedsin}).  Loma Prieta is chosen as an example of a more
recent sequence with a more complete record. It also exhibits
fluctuations in the local $p(t)$ greater than \( 2\sigma .\) Using a
running time window which is fixed to 0.1 days, the Omori's exponent
$p$ for the Westmoreland and for Loma Prieta aftershock sequences
appear to fluctuate log-periodically with a log frequency around 1.
The solid lines are for a constant number of events in the running
window, while the dashed lines are for fixed time windows.  In the
two examples shown here and for a majority of the real aftershock
sequences studied here, peaks are found at log frequencies near 1 (see
the peaks in solid and dashed lines close to the abscissa $1$ in
Figure \ref{windowkjaasdgaaqalo}).

\begin{figure}
  \psfrag{p-value}[][]{$p$ value}
  \psfrag{Time \(days\)}[ct][cB]{Time (days)}
  \figbox*{\hsize}{}{\includegraphics[width=8cm]{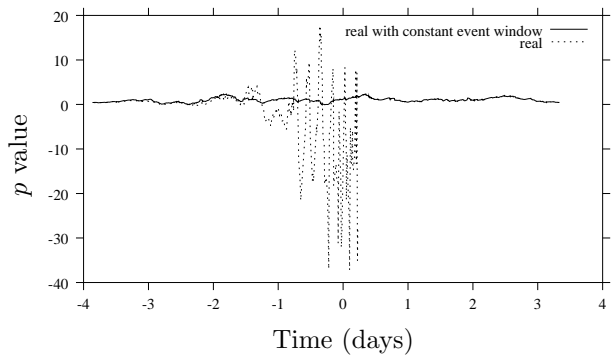}}
  \caption{Same as Figure \ref{event29fixedsin} for
    the Loma Prieta aftershock sequence.}
  \label{evenhfaasin}
\end{figure}
\begin{figure}
  \psfrag{Frequency}[ct][cB]{Frequency}
  \psfrag{Normalized Power}[][]{Normalized Lomb power}
  \figbox*{\hsize}{}{\includegraphics[width=8cm]{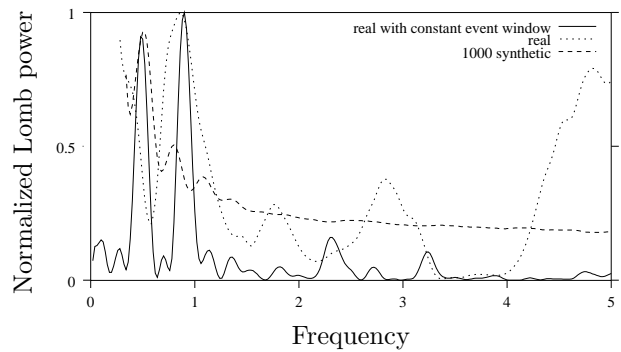}}
  \caption{Same as Figure \ref{windowkjaasdgaaqalo}
    for the Loma Prieta aftershock sequence. We also observe a common
    peak at a log frequency close to $1$ and a slightly smaller one at
    approximately half log frequency.}
  \label{winiiqalo}
\end{figure}

Table \ref{top3} lists the top three local peak maxima in frequency with
their associated spectral power from the Lomb periodogram for all
analyzable real aftershock sequences, using the maximum likelihood
method with fixed time window. The sequence numbers are as given by
\citet{kisslinger199107}. The amplitude of the fluctuations in
$p$ values in running time windows as a function of time is given in
 integer units of the standard deviation around
the average calculated for each sequence.

\begin{table*}
  \tablewidth{35pc}
  \caption{List of the Top Three Local Peak Maxima in Frequency With
Their Associated Spectral Power From the Lomb Periodogram for All
Analyzable Real Aftershock Sequences, Using the Maximum Likelihood
Method With Fixed Time Window\tablenotemark{\it a}\tablenotetext{\it
  a}{The sequence numbers are as \citet{kisslinger199107}. The
  amplitude of the fluctuations in $p$ values in running time windows
  as a function of time is given in integer units of the standard
  deviation around the average calculated for each sequence.}}
  \label{top3}
  \begin{center}
    \begin{tabular}{ccccccccc}\\[-3ex]\hline\\[-1ex]
Sequence&
Amplitude$\lp\sigma\rp$&
window (days)&
\( f_{1} \)&
\( f_{2} \)&
\( f_{3} \)&
\( power_{1} \)&
\( power_{2} \)&
\( power_{3} \)
      \\[1ex]\hline\\[-1ex]
1&
2&
150&
4.18&
3.61&
2.60&
13.74&
11.84&
8.42
\\
5&
1&
1&
0.71&
1.18&
1.65&
4.68&
2.52&
2.14
\\
6&
&
1&
1.61&
1.80&
1.38&
2.65&
2.58&
2.26
\\
7&
1&
100&
1.90&
0.79&
1.57&
5.67&
5.16&
5.11
\\
8&
1&
10&
0.56&
1.48&
4.20&
5.85&
3.42&
3.34
\\
9&
1&
1&
4.01&
4.95&
1.50&
2.83&
2.63&
2.61
\\
10&
&
1&
1.79&
1.47&
4.20&
3.13&
2.16&
1.75
\\
11&
&
1&
2.35&
2.63&
0.76&
2.00&
1.69&
1.53
\\
12&
&
1&
0.63&
3.14&
1.84&
4.27&
2.23&
1.48
\\
13&
1&
1&
1.48&
1.55&
2.48&
3.01&
2.95&
2.24
\\
14&
2&
100&
1.59&
3.65&
2.06&
16.73&
13.43&
5.49
\\
15&
&
1&
0.63&
0.89&
2.99&
3.37&
2.64&
2.54
\\
18&
&
1&
0.52&
4.39&
2.87&
4.57&
3.82&
3.78
\\
19&
&
10&
0.66&
1.33&
2.78&
3.92&
1.38&
1.31
\\
20&
&
1&
0.96&
2.82&
2.91&
3.91&
2.94&
2.86
\\
21&
&
10&
0.61&
1.22&
0.84&
2.18&
2.11&
1.60
\\
22&
1&
1&
2.47&
2.30&
0.69&
4.79&
4.69&
4.64
\\
23&
&
1&
1.19&
0.69&
1.88&
4.70&
1.38&
1.17
\\
24&
1&
1&
1.31&
1.07&
1.45&
4.96&
3.99&
3.66
\\
25&
1&
1&
2.53&
2.22&
1.99&
6.49&
4.97&
4.06
\\
26&
1&
1&
0.70&
4.13&
3.03&
22.82&
18.04&
11.46
\\
27&
&
1&
0.76&
4.34&
0.56&
2.56&
2.16&
2.01
\\
28&
2&
1&
1.25&
1.47&
2.36&
7.45&
7.07&
5.86
\\
29&
1&
0.1&
1.30&
3.84&
4.87&
13.07&
6.57&
6.08
\\
30&
&
1&
0.66&
3.55&
0.96&
5.61&
4.69&
3.07
\\
31&
2&
1&
4.06&
3.45&
1.72&
15.44&
5.87&
5.37
\\
32&
&
1&
0.53&
1.38&
1.68&
6.62&
5.12&
4.87
\\
33&
1&
1&
1.41&
3.57&
2.70&
30.14&
23.42&
16.79
\\
34&
2&
10&
0.58&
3.77&
1.50&
19.68&
15.43&
7.41
\\
35&
&
1&
0.55&
4.85&
1.65&
3.94&
3.08&
2.31
\\
36&
&
1&
1.42&
1.65&
2.22&
8.89&
8.42&
7.70
\\
Loma Prieta&
2&
0.1&
0.86&
4.83&
4.55&
15.91&
12.59&
9.53
\\
Landers&
2&
1&
4.06&
4.85&
0.97&
19.14&
17.11&
16.09
\\
Northridge&
2&
10&
0.60&
1.04&
4.20&
50.54&
49.38&
19.36
      \\[1ex] \hline\\[-3ex]
    \end{tabular}
  \end{center}
\end{table*}

\subsubsection{Lomb averaging.}

The case for
log-periodi-city in earthquake aftershocks is quite weak.
We have shown that the maximum likelihood method with running windows
of a fixed number of events generates spurious log-periodicity with a
log frequency inversely proportional to the window size. Furthermore,
the measured frequencies do not significantly deviate from those found
from 1000 synthetic sequences.  We moved to the maximum likelihood
method with a fixed time window to remove the log sampling effect. The
results were slightly more encouraging.  We have also tried to analyze
fluctuations of the difference between the cumulative number of
aftershocks and its smoothed Omori's law. Although the structures
generated by this method were in good agreement with the $p$ value
structure from the maximum likelihood method, the fact that both
methods give similarly correlated structures on synthetic data does
not enhance the evidence.

We now turn to a last investigation using averaging.  As shown by
\citet{johansen00:_punct}, averaging over Lomb periodograms provides
in principle a powerful method to retrieve the presence of
log-periodicity if the log frequencies are similar from sequence to
sequence. This is due to the fact that the Lomb periodograms are not
sensitive to the phase of the log-periodic oscillations, and we do not
run the risk of scrambling the log-periodic oscillations due to their
random phases, as would occur during a direct averaging of the $p(t)$
functions for instance.  See also \citet{johansen199805} for a general
discussion of the caveats of ensemble averaging.

\begin{figure}
  \psfrag{Frequency}[ct][cB]{Frequency}
  \psfrag{Normalized Power}[][]{Normalized Lomb power}
  \figbox*{\hsize}{}{\includegraphics[width=8cm]{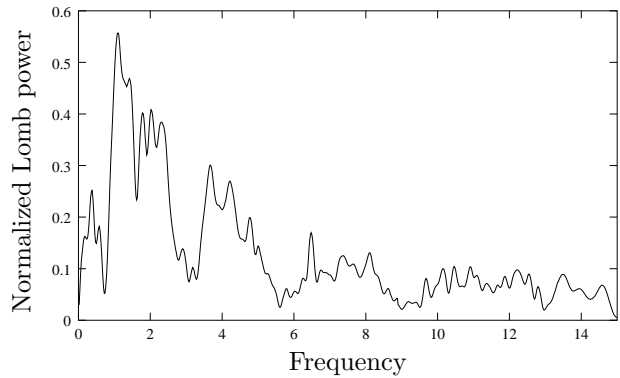}}
  \caption{Average Lomb periodogram over all real aftershock
    sequences using the maximum likelihood method with running windows
    of a constant number of events (log sampling).}
  \label{evesin}
\end{figure}

\begin{figure}
  \psfrag{Frequency}[ct][cB]{Frequency}
  \psfrag{Normalized Power}[][]{Normalized Lomb power}
  \figbox*{\hsize}{}{\includegraphics[width=8cm]{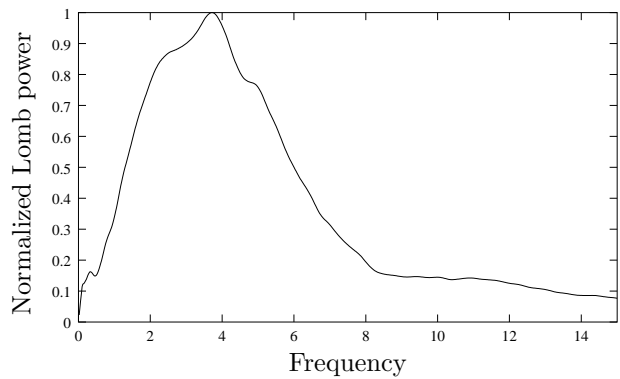}}
  \caption{Same as Figure \ref{evesin} but for 1000
    synthetic sequences for each real aftershock sequence with the
    same parameters.}
  \label{wiiiaaqalo}
\end{figure}

Figure \ref{evesin} shows the average Lomb periodogram obtained by
using all real aftershock sequences. It should be compared to Figure
\ref{wiiiaaqalo} obtained by averaging the Lomb periodograms of 1000
synthetic realizations for each real aftershock sequence. We see that
the peak is differently positioned for the real aftershock sequences
compared to that of the synthetic.


\begin{figure}
  \psfrag{Frequency}[ct][cB]{Frequency}
  \psfrag{Normalized Power}[][]{Normalized Lomb power}
  \figbox*{\hsize}{}{\includegraphics[width=8cm]{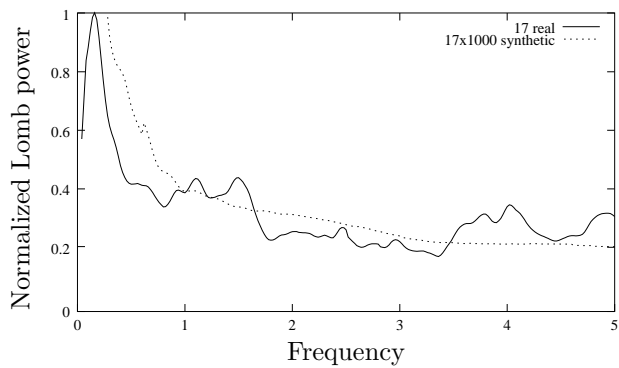}}
  \caption{The solid line is the
    average Lomb periodogram over the subset of 17 real aftershock
    sequences for which the $p$ value fluctuations are greater than $1
    \sigma$. The dashed line is the average over $17,000$ synthetic
    aftershock sequences, 1000 for each of the 17 real ones with
    $p$ value fluctuations greater than $1 \sigma$.}
  \label{evesinaad}
\end{figure}

\begin{figure}
  \psfrag{Frequency}[ct][cB]{Frequency}
  \psfrag{Normalized Power}[][]{Normalized Lomb power}
  \figbox*{\hsize}{}{\includegraphics[width=8cm]{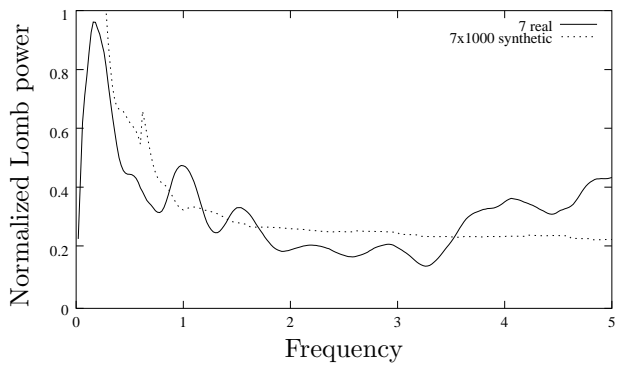}}
  \caption{The solid line is the average Lomb periodogram over
    the subset of seven real aftershock sequences for which the $p$ value
    fluctuations are greater than $2 \sigma$.  The dashed line is the
    average over 7000 synthetic aftershock sequences, 1000 for each of
    the seven real ones with $p$ value fluctuations greater than $2
    \sigma$.}
  \label{wiiadaqiaaqalo}
\end{figure}

Encouraged by this, we have singled out all real aftershock sequences
where the fluctuations around the average Omori's law of $p=1$ was
greater than $1$ and $2$ $\sigma$. The result is shown in Figures
\ref{evesinaad} and \ref{wiiadaqiaaqalo}.  We see that a stronger
log-periodic signal appears in the spectra of the real aftershock
sequences compared to that of the synthetic. By this, we observe that
the peak culminating at 1 of the solid line (real data) is
significantly sharper than that for the synthetic ensemble.
This indicates that genuine log-periodic structures might be most
easily identified in sequences with ``anomalous'' high or low
$p$ value and large fluctuations.


\section{Discussion}
\label{sec:discussion}

We have presented a detailed study of log-periodicity in relatively
sparse data with nonuniform sampling. We have focused our attention
on the effect of this nonuniform sampling and its interplay with
integration in the presence of noise, which may lead to synthetically
generated log-periodicity that may compete with genuine signals.  Our
main result is the discovery that approximately logarithmic sampling
and integration (performed either by constructing cumulative functions
or using a maximum likelihood method) generate strong log-periodic
structures.  This result calls for a reassessment of previous reports
of log-periodicity in rupture
\citep{anifrani199506,sahimi9610,huang199706,johansen199805} and in
earthquakes
\citep{sornette199505,saleur199603,saleur199603a,saleur199608,johansen199610,varnes199601,leephd}
in the spirit of the reanalysis \citep{johansen99:_new_kobe} of the
evidence for criticality and especially log-periodicity in the
previously reported chemical anomalies that preceded the Kobe
earthquake. In this recent work \citep{johansen99:_new_kobe}, the
ion (Cl$^-$, K$^+$, Mg$^{++}$, NO$_3^{-}$ and SO$_4^{--}$)
concentrations of groundwater issued from deep wells located near the
epicenter of the 1995 Kobe earthquake are taken as proxies for the
cumulative damage preceding the earthquake. Using both a parametric
and nonparametric analysis following the guidelines given in this
 paper, the five data sets are compared extensively to
synthetic time series.  The null hypothesis that the patterns
documented on these times series result from noise decorating a simple
power law is rejected with a very high confidence level. For the other
cases of reported log-periodicity, we plan to present a reanalysis in
forthcoming publications.

We have also analyzed in detail the 27 best aftershock sequences
studied by \citet{kisslinger199107} and the Loma Prieta, Landers, and
Northridge aftershock sequences for the presence of log-periodic
corrections to Omori's law. Both the cumulative distribution and
maximum likelihood methods lead to synthetic log-periodicity resulting
from the interplay between noise reddening and power law scaling.
However, it is important to distinguish between the two processes.
The low-pass filtering resulting from the integration used to generate
the cumulative distribution is very different from the averaging
process used in the maximum likelihood method. In the latter, the
upper frequency cutoff is completely determined by the data window
used in the estimation process. Most of the structures observed on the
27 aftershock sequences can be explained solely on the basis of the
spurious log-periodicity generated by the nonuniform sampling and the
integration mechanism. There are some residual structures in
aftershock sequences of possible log-periodic origin which cannot be
fully explained by the synthetic scenario. Hence the possibility of a
signal resulting from a physically based log-periodicity as claimed by
\citet{leephd} cannot be ruled out.  Better quality data with an order
of magnitude more events are needed to resolve this issue.

A tantalizing question is whether the spurious log-periodic
oscillations that might appear in the analysis of a foreshock
sequence can still be used to improve the prediction of the main
event. This would be equivalent to ``fitting the most probable form of
noise,'' an attractive though maybe not workable idea.  We hope to
report on this point soon.

\acknowledgments 
We are grateful to C. Kisslinger for sharing his data with us.

\balance 

\bibliographystyle{agu}
\bibliography{yueqiangRef}

\end{article}
\end{document}